\documentclass[pra,showpacs,showkeys,superscriptaddress,amssymb,amsmath]{revtex4}
\usepackage{graphicx}
\usepackage{hyperref}
\usepackage{placeins}
\usepackage{xcolor}
\usepackage{caption}
\usepackage{listings}
\usepackage[normalem]{ulem}
\usepackage{color}

         \newcommand{\correct}[1]{{\color{red}#1}}

\begin{document}

	\title{
		A combinatorial matrix approach for the generation of vacuum Feynman graphs multiplicities in $\phi^{4}$ theory
	}
	
	\author{Erick Castro}
	\email[]{erickc@cbpf.br}
	\affiliation{Centro Brasileiro de Pesquisas F\'{\i}sicas/MCTI,
		22290-180, Rio de Janeiro, RJ, Brazil}	
	
	\author{I. Roditi}
	\email[]{roditi@cbpf.br}
	\affiliation{Centro Brasileiro de Pesquisas F\'{\i}sicas/MCTI,
		22290-180, Rio de Janeiro, RJ, Brazil}	
	
	\begin{abstract}
		From the standard procedure for constructing Feynman vacuum graphs in $\phi^{4}$ theory from the generating functional $Z$, we find a relation with sets of certain combinatorial matrices, which allows us to generate the set of all Feynman graphs and the respective multiplicities in an equivalent combinatoric way. These combinatorial matrices are explicitly related with the permutation group, which facilitates the construction of the vacuum Feynman graphs. Various insights in this combinatoric problem are proposed, which in principle provide an efficient way to compute Feynman vacuum graphs and their multiplicities. 
	\end{abstract}
	
	\maketitle
	
	\section{Introduction \label{Int}}
	
	The problem of generating Feynman graphs and their respective multiplicities is recurrent, both in Quantum Field Theory (QFT) and many body solid state (MB) literature. In the functional approach, the starting point of such constructions is the generating functional of the correlation functions $Z$ and the associate Legendre transformations, which generates specific types of Feynman graphs (connected and one-particle irreducible, see ref.\cite{Justin} for a concise introduction). The paradigmatic case in order to introduce Feynman diagrammatic machinery in QFT is generally $\phi^{4}$ theory which lies at the heart of second-order phase-transition phenomena. Diagrammatic counting techniques are no exception, one instance is the use of zero-dimensional $\phi^{N}$ theory to generate the number of total Feynman graphs in each perturbation order \cite{Cvitanovic} (in particular, connected vacuum diagrams and connected graphs with external legs), extending then to QED and non-abelian field theories. In the MB context the use of Feynman graphs techniques is extensive, in many cases the quasi-particle principle is applicable and expressed mathematically in term of self-consistent approximations, for example Hedin's system of equations \cite{Hedin}, which involve physical process expressed by Feynman diagrams. In  ref.\cite{Molinari}, Hedin's system of equations is treated in zero-dimension leading to the explicit counting of the Feynman graphs that contribute in this quasi-particle regimen. Ref.\cite{Pavlyukh1} extends this treatment and shows interesting relations among QFT and combinatorics of Feynman graphs. 
	
	In the construction of the Feynman graphs by functional derivatives (or equivalently, by contractions in the field operator approach, see for example ref.\cite{Castro}) some diagrams generated by different derivatives (or different contractions) are equivalent, this appears as a multiplicity of equivalent diagrams. Such multiplicities in a given diagram $\mathcal{G}$, are related to the symmetry factor of  $\mathcal{G}$, which
	is in connection with the number of automorphisms $\left| \mathrm{Aut}(\mathcal{G})\right| $ of the graph, see  ref.\cite{arXiv1} and ref.\cite{Dong} for example. The determination of $\left| \mathrm{Aut}(\mathcal{G})\right| $  for a given graph $\mathcal{G}$ is in general a difficult problem \cite{Beals}, therefore, the same can be said with respect to the multiplicities of $\mathcal{G}$. Nonetheless, this does not prevent the existence of practical computational algorithms for the generation of Feynman graphs \cite{Nogueira}. 
	
	In a global sense, the complete information of the multiplicities is contained in the generating fuctional of the Feynman graphs. This is evident in the zero dimensional approach. In this case, for a given order, all the diagrams are equivalent and the number of Feynman graphs appear explicitly multiplying the coupling constant. In the general case, the construction of Feynman diagrams starting from the generating functional is an arduous task for growing perturbation orders, since the number of functional derivatives increase rapidly. Nevertheless, some achievements have been reached in this respect: ref.\cite{Kleinert} introduce differential functional calculus to get recursive relations, which, iteratively determine the multiplicities for connected vacuum diagrams. In this reference it is shown that the $m$-order vacuum connected diagrams determine the other connected $m$-order diagrams (Feynman graphs with external legs), and therefore the corresponding multiplicities in a simple way.
	
	On the other hand, relations between combinatorics and perturbative quantum field theory have been established gradually, for example, in the nature of renormalization group structure \cite{Connes} \cite{Kreimer} or with problems in combinatorial theory \cite{Nakanishi}. With respect to the problem of counting of objects in quantum field theories, interesting relationships with the permutation group and string theories have been established (see for example ref.\cite{Koch}). In this respect, interesting correspondences between N-rooted maps and Feynman diagrams in QED field theory have been put in evidence, see Ref.\cite{arXiv2}. In the recent reference \cite{Brádler} the calculation of multiplicities for bosonic (scalar) theories is put in correspondence with the solutions of certain system of Diophantine equations. 
	
	In this work, we explore a different route. From the generating functional $Z$ in $\phi^4$ theory, we establish subset classes of $m$-order vacuum Feynman graphs (connected and disconnected) for each perturbative order $m$. Every diagram in each subset have associated a natural number. When adding all the numbers associated to the same diagram in all the subsets we obtain the corresponding multiplicity. This subsets of Feynman graphs are indexed by the equivalence classes in the total set of certain types of $m\times m$ matrices with equal row and column sums (RC-magic squares). The equivalence classes are induced naturally by the permutation group. In particular, every RC-magic square in the equivalence class determine the associated subset of vacuum diagrams in a simple way. Thus, the problem of constructing vacuum Feynman graphs is reduced to finding one representative square matrix in each equivalence class of the total RC-magic square set. The relation between Feynman diagrams and RC-magic squares is found directly from the perturbative expansion of the generating functional $Z$.  We develop a straightforward algorithm for the generation of the RC-magic squares representatives which are inequivalent matrices belonging to different equivalent classes. Based on the relationship between the total set of RC-magic squares and the well-known Birkhof-von Neumann theorem, we find a criterion to distinguish two inequivalent matrices, which could simplify the computation of the RC-magic square representatives and therefore the Feynman graphs multiplicities. RC-magic squares are interesting combinatorial objects, the matrices studied here are simpler than those known by Chinese, Indian and Arab civilizations which had the same row, column and diagonal components sum. Here, our RC-magic squares only have identical row and column sum. For matrix combinatorial theory bibliography see for example ref.\cite{Brualdi}. 
	
	This paper is organized as follows. In sec.\ref{sec2} we unambiguously establish  the relation between certain subset of $m$-order vacuum  Feynman graphs and the total set of $m\times m$ RC-magic squares with sum equal to 4. The subsets are generated directly from the expansion of the functional generator $Z$ and the relation with the RC-magic squares is explicitly demonstrated. The relation is formulated within the adjacency matrix notation for graph theory, which facilitates the computation of the Feynman graphs associated with the respective RC-magic square. In Sec.\ref{sec3} we prove that the action of the permutation group, as an  interchange between rows, and an interchange between columns of a specific RC-magic square does not change the associated subset of the Feynman vacuum graphs. This fact, induces in the total $m\times m$ RC-magic square set an equivalence relation, whose equivalence classes are formed by equivalent matrices that generates the same subset of $m$-order Feynman graphs. As an example, we construct all the third-order Feynman graphs using only one representative in each equivalence class in the total set of the $3\times 3$ RC-magic squares. Finally, in sec.\ref{sec5} we discuss the efficiency of an algorithm, whose code is implemented in the appendix \ref{IV}, for the calculation of the RC-magic square representatives and the size (number of elements) of each equivalence class. We put in evidence certain properties of the permutation matrices, which together with the Birkhoff-von Neumann theorem could allow some simplification on the construction of the RC-magic squares representatives. Sec.\ref{sec6} contains discussion and conclusions.     
	
	\section{Relation between the expansion of the generating functional and RC-magic squares}\label{sec2}    
	  
	To avoid the appearance of the imaginary unit, we start our study with the generating functional $Z$ in euclidean $\phi^4$ theory, coupled with an external field $J(x)$, $x \in \mathbb{R}^{d}$. According to the standard procedure \cite{Justin}, we have $Z(J)$ in the formal perturbative form:
	
	\begin{equation}
	Z(J)=\exp\left[-\mathcal{V}\left(\frac{\delta}{\delta J}\right)\right]\exp\left(\frac{1}{2}J\Delta J\right)\label{Partition},
	\end{equation} 
	with $$\frac{1}{2}J\Delta J \rightarrow \frac{1}{2}\int \mathrm{d}^{d}y\, \mathrm{d}^{d}z \,J(y)\Delta(y,z)J(z),$$ and $$ \mathcal{V}\left(\frac{\delta}{\delta J}\right) \rightarrow \frac{g}{4!}\int \mathrm{d}^{d}x\frac{\delta^{4}}{\delta J^{4}(x)},$$ where $\Delta(y,z)$ is the free propagator and $g$ the coupling constant.
	
	The above expressions make sense only perturbatively. Thus, expanding the two exponentials and taking the formal limit $J \to 0$, $Z(J\to 0)$ generates the vacuum Feynman diagrams. Particularly, the $m$-order diagrams are in the expression
	
	\begin{align}
	\frac{1}{(2m)!2^{2m}}&\frac{g^{m}}{(4!)^{m}m!}\int \mathrm{d}^{d}x_{1}\cdots \mathrm{d}^{d}x_{m}\,\frac{\delta^{4}}{\delta J^{4}(x_{1})}\cdots \nonumber \\ \times&\frac{\delta^{4}}{\delta J^{4}(x_{m})}\left[\left(J(y)\Delta(y,z)J(z)\right)^{2m}\right]. \label{Der}
	\end{align}
	
	Note that the number of functional derivatives in the above expression is equal to the number of $J$'s in the derivative argument, therefore, this is the only term that survives for all $m$ when $J \to 0$. We define 
	
	\begin{equation}
	B(J,J,J,J)=\left(J\Delta J\right)^{2}
	\end{equation}
	applying the chain rule in (\ref{Der}) over the $B$ products and adding the terms that generate the same function, we obtain a number $\mathcal{N}(m)$ of summing terms. Over each $B$ in these terms act four functional derivatives. Assuming $\Delta(x,y)=\Delta(y,x)$ and defining
	
	\begin{equation}\label{DerB}
	\frac{\delta}{\delta J(x)}\frac{\delta}{\delta J(y)}\frac{\delta}{\delta J(z)}\frac{\delta}{\delta J(w)}\left[B\right]\equiv \aleph(x,y,z,w)
	\end{equation}
	we have
	
	\begin{align}
	\aleph(x,y,z,w)=& 8\Delta(x,y)\Delta(z,w)\nonumber\\+8&\Delta(x,z)\Delta(y,w)+8\Delta(x,w)\Delta(y,z). \label{aleph}
	\end{align}
	
	In particular we have
	
	\begin{align}
	\aleph(x,x,y,y)&=8\Delta(x,x)\Delta(y,y)+16\Delta^{2}(x,y)\\ 
	\aleph(x,x,y,z)&=8\Delta(x,x)\Delta(y,z)+16\Delta(x,y)\Delta(x,z)\\	
	\aleph(x,x,x,y)&=24\Delta(x,x)\Delta(x,y)\\
	\aleph(x,x,x,x)&=24\Delta^{2}(x,x). \label{Inf}
	\end{align}
	
	If we change the order of the variables in (\ref{aleph}), we generate the same function. Now, each one of the $\mathcal{N}(m)$ summing terms have $m$ products of $\aleph$'s functions. Let us think about the following problem, be a set of $4m$ balls of $m$ different colors such that, for each color, we have 4 balls. Be a box with $m$ compartments such that every compartment only supports 4 balls. In how many different ways can these $4m$ balls  be distributed in the box? It is evident that (\ref{Der}) is also a generating function of this combinatorial problem; there are $4m$ variables introduced by the $4m$ functional derivatives, the $m$ variables $\{x_{1},x_{2},\cdots,x_{m}\}$ correspond to the $m$ colors, every $\aleph$ is a compartment with four variables and the chain rule of the derivatives, after adding identical contributions, gives all the possibilities of distribution. Therefore, there are $\mathcal{N}(m)$ ways to distribute the $4m$ balls in the box.
	
	An arbitrary configuration in the $4m$ distribution problem can be represented by a matrix (see TABLE \ref{Table1}). The $a_{ij}$ components are the number of $j$-color balls presents in the compartment $i$, evidently $a_{ij} \in \{0,1,2,3,4\}$ for all $i$ and $j$. Every compartment has $4$ balls, therefore, we have $$\sum_{j=1}^{m}a_{ij}=4, $$ for each $i$. In the same way, there exist $4$ balls of each color, then $$\sum_{i=1}^{m}a_{ij}=4$$ for each $j$. This means that each term of (\ref{Der}) in the $\aleph$'s multiplicative form can be represented by a matrix with equal row and column sum (RC-magic square). 
	
\begin{center}
	\captionof{table}{An arbitrary configuration in the combinatorial $4m$ balls distribution problem.}
	\label{Table1}
	\begin{tabular}{ c | c |c |c | c }
		
		       & color 1 & color 2 & $\cdots$ & color $m$  \\ \hline
		Comp 1 & $a_{11}$       & $a_{12}$       &  $\cdots$&      $a_{1m}$      \\ \hline
		Comp 2 & $a_{21}$       & $a_{22}$       &  $\cdots$&      $a_{2m}$      \\ \hline
	   $\vdots$& $\vdots$       & $\vdots$      &  $\vdots$&     $\vdots$      \\ \hline
	    Comp m & $a_{m1}$       & $a_{m2}$       &  $\cdots$&    $a_{mm}$     \\ \hline
		\hline
	\end{tabular}
\end{center}
	
	 An example for $m=4$:
	 
	\begin{equation}\label{Matrix}
	\left(\begin{array}{cccc}
	1 & 1 & 1 & 1 \\
	3 & 1 & 0 & 0 \\
	0 & 2 & 1 & 1 \\
	0 & 0 & 2 & 2 
	\end{array}\right)\to \begin{array}{c}
	\aleph(x_{1},x_{2},x_{3},x_{4})\times\aleph(x_{1},x_{1},x_{1},x_{2})\\\times\aleph(x_{2},x_{2},x_{3},x_{4})\times\aleph(x_{3},x_{3},x_{4},x_{4}) \end{array}
	\end{equation}
	
This relation is implicit using the Table I, substituting the compartments by the $\aleph$ functions, and the color $i$ by the variable $x_{i}$. In our particular case, TABLE \ref{Table2} exemplifies
 the association between RC-magic square and the $\aleph$'s product in (\ref{Matrix}).

	\begin{center}
	\captionof{table}{The construction of the $\aleph$ product associated with the RC-magic square in (\ref{Matrix}).}
	\label{Table2}
	\begin{tabular}{ c | c |c |c | c }
		
		       & $x_{1}$ & $x_{2}$ & $x_{3}$ &  $x_{4}$  \\ \hline
 First $\aleph$ & $1$       & $1$   &  $1$&      $1$      \\ \hline
 Second $\aleph$ & $3$       & $1$       &  $0$&      $0$      \\ \hline
Third $\aleph$& $0$       & $2$      &  $1$&     $1$      \\ \hline
Fourth $\aleph$& $0$       & $0$       &  $2$&    $2$     \\ \hline
		\hline
	\end{tabular}
\end{center}
	
	As we mentioned before, the chain rule in (\ref{Der}) contains all the possibles ways for constructing a RC-magic square. Some ways produce identical results, for example in (\ref{DerB}) there are $4!$ ways of constructing $\aleph(x,y,z,w)$. In the general case, for an arbitrary RC-magic square $m\times m$, the multiplicity is the following product
	
	\begin{equation}\label{Prod}
	\prod_{i=1}^{m}\left[\frac{4!}{a_{i1}!a_{i2}!\cdots a_{im}!}\right],
	\end{equation}
for (\ref{Matrix}), we have the multiplicity $24\times 4\times 12\times 6=6912$.
	
	Now, for any RC-magic square the next question arises: what are the associated Feynman graphs? For a $m\times m$ matrix is necessary to multiply the $m$ $\aleph$'s functions associated, this generates a sum of free propagators products, which correspond to the Feynman graphs. There is an algebraic way to implement this, each $\aleph$ function have at most four $x_{i}$ different variables, with $1 \leq i \leq m $, which are the vertices of a certain sum of propagators product terms given by one of the equations (\ref{aleph})-(\ref{Inf}); every one of this terms is a graph, which can be represented by an $n\times n$ adjacency matrix (with $n$ the different vertices of the graph, thus $1\leq n \leq 4$). Given $\{x_{a},x_{b},x_{c},x_{d}\}$ with $1\leq a < b < c < d \leq m$, the possible adjacency matrices are

\begin{equation}\label{12}
\aleph(x_{a},x_{b},x_{c},x_{d}) \to 8\left[\begin{array}{cccc}
0 & 1 & 0 & 0 \\
1 & 0 & 0 & 0 \\
0 & 0 & 0 & 1 \\
0 & 0 & 1 & 0 
\end{array}\right]
+8\left[\begin{array}{cccc}
0 & 0 & 1 & 0 \\
0 & 0 & 0 & 1 \\
1 & 0 & 0 & 0 \\
0 & 1 & 0 & 0 
\end{array}\right]
+8\left[\begin{array}{cccc}
0 & 0 & 0 & 1 \\
0 & 0 & 1 & 0 \\
0 & 1 & 0 & 0 \\
1 & 0 & 0 & 0 
\end{array}\right]
\end{equation}

\begin{equation}\label{13}
\aleph(x_{a},x_{a},x_{b},x_{b}) \to 8\left[\begin{array}{cc}
2 & 0  \\
0 & 2   
\end{array}\right]
+16\left[\begin{array}{cc}
0 & 2 \\
2 & 0  
\end{array}\right]
\end{equation}

\begin{equation}
\aleph(x_{a},x_{a},x_{b},x_{c}) \to 8\left[\begin{array}{ccc}
2 & 0 & 0 \\
0 & 0 & 1 \\
0 & 1 & 0 
\end{array}\right]
+16\left[\begin{array}{ccc}
0 & 1 & 1 \\
1 & 0 & 0 \\
1 & 0 & 0 
\end{array}\right]
\end{equation}

\begin{equation}
\aleph(x_{a},x_{b},x_{b},x_{c}) \to 8\left[\begin{array}{ccc}
0 & 0 & 1 \\
0 & 2 & 0 \\
1 & 0 & 0 
\end{array}\right]
+16\left[\begin{array}{ccc}
0 & 1 & 0 \\
1 & 0 & 1 \\
0 & 1 & 0 
\end{array}\right]
\end{equation}	

\begin{equation}
\aleph(x_{a},x_{b},x_{c},x_{c}) \to 8\left[\begin{array}{ccc}
0 & 1 & 0 \\
1 & 0 & 0 \\
0 & 0 & 2 
\end{array}\right]
+16\left[\begin{array}{ccc}
0 & 0 & 1 \\
0 & 0 & 1 \\
1 & 1 & 0 
\end{array}\right]
\end{equation}	

\begin{equation}
\aleph(x_{a},x_{a},x_{a},x_{b}) \to 24\left[\begin{array}{cc}
2 & 1 \\
1 & 0 
\end{array}\right]
\end{equation}

\begin{equation}\label{15}
\aleph(x_{a},x_{b},x_{b},x_{b}) \to 24\left[\begin{array}{cc}
0 & 1 \\
1 & 2 
\end{array}\right]
\end{equation}
	
\begin{equation}\label{16}
\aleph(x_{a},x_{a},x_{a},x_{a}) \to 24\left[\begin{array}{c}
4 
\end{array}\right]
\end{equation}

Note that the usual order in the indices of the variables $\{x_{a},x_{b},x_{c},x_{d}\}$ was chosen, this only to facilitate the construction of the adjacency matrices. The notation with brackets in the matrices means that we are working with adjacency matrices. Adjacency matrices are always symmetric square matrices (do not confuse with the RC-magic square set, despite the fact that each adjacency matrix is by coincidence a symmetric RC-magic square) whose coefficients represents the number of lines (edges) joining two vertices. So, $a_{ij}=a_{ji}$ is the number of lines joining the vertices $x_{i}$ and $x_{j}$. If $j=i$ the number of lines of $x_{i}$ in itself (loops) is usually multiplied by 2, therefore the diagonals coefficients are even, see ref.\cite{Diestel}. In ref.\cite{Kleinert} the adjacency matrix representation is also used.

 We can think of (\ref{12})-(\ref{16}) as the 8 basic blocks to build all the Feynman graphs, namely, when we multiply the $\aleph$'s functions of a RC-magic square, the distributive multiplication property offers all the allowed ways to ``assemble" the blocks. How we can interpreted this process in the adjacency matrix notation? Every one of this blocks have one $m \times m$ matrix representation, for example, to $\aleph(x_{2},x_{2},x_{4},x_{4})$ the $5\times 5$ representation of (\ref{13}) for $x_{a}=x_{2}$ and $x_{b}=x_{4}$ is

\begin{equation}
\aleph(x_{2},x_{2},x_{4},x_{4}) \to 8\left[\begin{array}{ccccc}
0 & 0 & 0 & 0 & 0 \\
0 & 2 & 0 & 0 & 0 \\
0 & 0 & 0 & 0 & 0 \\
0 & 0 & 0 & 2 & 0 \\
0 & 0 & 0 & 0 & 0
\end{array}\right]
+16\left[\begin{array}{ccccc}
0 & 0 & 0 & 0 & 0 \\
0 & 0 & 0 & 2 & 0 \\
0 & 0 & 0 & 0 & 0 \\
0 & 2 & 0 & 0 & 0 \\
0 & 0 & 0 & 0 & 0
\end{array}\right]
\end{equation}
or, the $3 \times 3$ representation of (\ref{16}) for $x_{a}=x_{2}$ is

\begin{equation}
\aleph(x_{2},x_{2},x_{2},x_{2}) \to 24\left[\begin{array}{ccc}
0 & 0 & 0 \\
0 & 4 & 0 \\
0 & 0 & 0 
\end{array}\right],
\end{equation}
when we multiply two $\aleph$'s functions, the corresponding products of propagators (\ref{aleph})-(\ref{Inf}) are multiplied in every possible way (according to the distributive property) and this is equivalent to add the associated adjacency matrices in the $m\times m$ representation. This can be computationally implemented as follows: instead of replacing the propagator products by the associated $m\times m$ adjacency matrices, we replace each one of them by a formal function $E(M)$ whose argument $M$ is the correspondent $m\times m$ adjacency matrix. This formal definition $E(\cdots)$ also satisfies \begin{equation}E(M_{1})\times E(M_{2})=E(M_{1}+M_{2}).\label{Mult}\end{equation} This multiplicative property guarantees that the $m\times m$ adjacency matrices belonging to different $\aleph$'s be added when we perform the $\aleph$'s product. At the end, we get a sum of $E$ functions whose arguments are the adjacency matrices of the searched $m$-order Feynman graph. As an example, let us look at the fourth order Feynman graphs associated with (\ref{Matrix}) in the adjacency matrix $4\times 4$ representation:

	\begin{align}
			\left(\begin{array}{cccc}
				1 & 1 & 1 & 1 \\
				3 & 1 & 0 & 0 \\
				0 & 2 & 1 & 1 \\
				0 & 0 & 2 & 2 
			\end{array}\right) \to& \left\{8E\left(\left[\begin{array}{cccc}
			0 & 1 & 0 & 0 \\
			1 & 0 & 0 & 0 \\
			0 & 0 & 0 & 1 \\
			0 & 0 & 1 & 0 
		\end{array}\right]\right)+8E\left(\left[\begin{array}{cccc}
		0 & 0 & 1 & 0 \\
		0 & 0 & 0 & 1 \\
		1 & 0 & 0 & 0 \\
		0 & 1 & 0 & 0 
	\end{array}\right]\right)+8E\left(\left[\begin{array}{cccc}
	0 & 0 & 0 & 1 \\
	0 & 0 & 1 & 0 \\
	0 & 1 & 0 & 0 \\
	1 & 0 & 0 & 0 
\end{array}\right]\right)\right\}\times \left\{24E\left(\left[\begin{array}{cccc}
2 & 1 & 0 & 0 \\
1 & 0 & 0 & 0 \\
0 & 0 & 0 & 0 \\
0 & 0 & 0 & 0 
\end{array}\right]\right)\right\}\nonumber \\ \times&\left\{8E\left(\left[\begin{array}{cccc}
0 & 0 & 0 & 0 \\
0 & 2 & 0 & 0 \\
0 & 0 & 0 & 1 \\
0 & 0 & 1 & 0 
\end{array}\right]\right)+16E\left(\left[\begin{array}{cccc}
0 & 0 & 0 & 0 \\
0 & 0 & 1 & 1 \\
0 & 1 & 0 & 0 \\
0 & 1 & 0 & 0 
\end{array}\right]\right)\right\}\times \left\{8E\left(\left[\begin{array}{cccc}
0 & 0 & 0 & 0 \\
0 & 0 & 0 & 0 \\
0 & 0 & 2 & 0 \\
0 & 0 & 0 & 2 
\end{array}\right]\right)+16E\left(\left[\begin{array}{cccc}
0 & 0 & 0 & 0 \\
0 & 0 & 0 & 0 \\
0 & 0 & 0 & 2 \\
0 & 0 & 2 & 0 
\end{array}\right]\right)\right\} \label{FeynD1}
	\end{align}
	
 After applying the distributive property we have a sum of 12 $E$ functions whose arguments are 12 adjacency matrices which generate the associated 4-order Feynman graphs. Some of this adjacency matrices generates the same Feynman graphs. For instance one of this twelve terms is:

	\begin{align}
			 \to& \left\{8E\left(\left[\begin{array}{cccc}
			0 & 1 & 0 & 0 \\
			1 & 0 & 0 & 0 \\
			0 & 0 & 0 & 1 \\
			0 & 0 & 1 & 0 
		\end{array}\right]\right)
		\right\}\times \left\{24E\left(\left[\begin{array}{cccc}
2 & 1 & 0 & 0 \\
1 & 0 & 0 & 0 \\
0 & 0 & 0 & 0 \\
0 & 0 & 0 & 0 
\end{array}\right]\right)\right\}\times\left\{8E\left(\left[\begin{array}{cccc}
0 & 0 & 0 & 0 \\
0 & 2 & 0 & 0 \\
0 & 0 & 0 & 1 \\
0 & 0 & 1 & 0 
\end{array}\right]\right)
\right\}\times \left\{8E\left(\left[\begin{array}{cccc}
0 & 0 & 0 & 0 \\
0 & 0 & 0 & 0 \\
0 & 0 & 2 & 0 \\
0 & 0 & 0 & 2 
\end{array}\right]\right)\right\} \label{FeynDEx1}
	\end{align}

	\begin{align}
			 \to 8\times24\times8\times8E\left(\left[\begin{array}{cccc}
			0 & 1 & 0 & 0 \\
			1 & 0 & 0 & 0 \\
			0 & 0 & 0 & 1 \\
			0 & 0 & 1 & 0 
		\end{array}\right]+
		\left[\begin{array}{cccc}
2 & 1 & 0 & 0 \\
1 & 0 & 0 & 0 \\
0 & 0 & 0 & 0 \\
0 & 0 & 0 & 0 
\end{array}\right]+\left[\begin{array}{cccc}
0 & 0 & 0 & 0 \\
0 & 2 & 0 & 0 \\
0 & 0 & 0 & 1 \\
0 & 0 & 1 & 0 
\end{array}\right]+
\left[\begin{array}{cccc}
0 & 0 & 0 & 0 \\
0 & 0 & 0 & 0 \\
0 & 0 & 2 & 0 \\
0 & 0 & 0 & 2 
\end{array}\right]\right) \label{FeynDEx2}
	\end{align}
		
	\begin{align}
			 \to 12288E\left(\left[\begin{array}{cccc}
			2 & 2 & 0 & 0 \\
			2 & 2 & 0 & 0 \\
			0 & 0 & 2 & 2 \\
			0 & 0 & 2 & 2 
		\end{array}\right]\right) \to \underbrace{\includegraphics[scale=0.2]{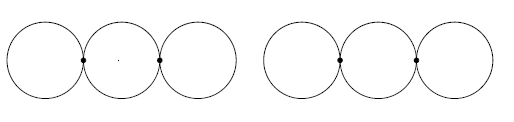}}_{12288}\label{FeynDEx3}
	\end{align}
	
In particular we have 7 differents Feynman graphs with the respective multiplicities:

 \begin{align}
 	\left(\begin{array}{cccc}
 	1 & 1 & 1 & 1 \\
 	3 & 1 & 0 & 0 \\
 	0 & 2 & 1 & 1 \\
 	0 & 0 & 2 & 2 
 	\end{array}\right)\to \underbrace{\includegraphics[scale=0.2]{diag1.png}}_{12288}+ \underbrace{\includegraphics[scale=0.2]{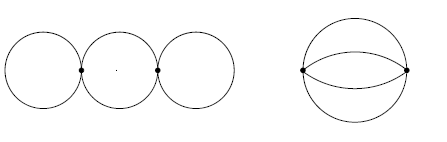}}_{24576}+\underbrace{\includegraphics[scale=0.12]{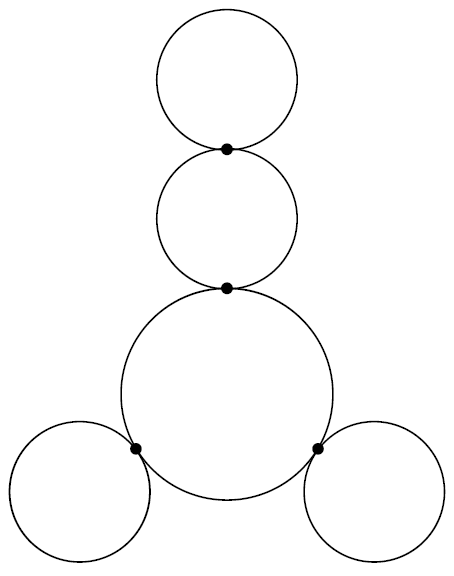}}_{73728}+\underbrace{\includegraphics[scale=0.12]{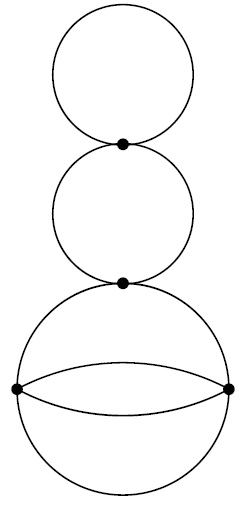}}_{49152}+\underbrace{\includegraphics[scale=0.16]{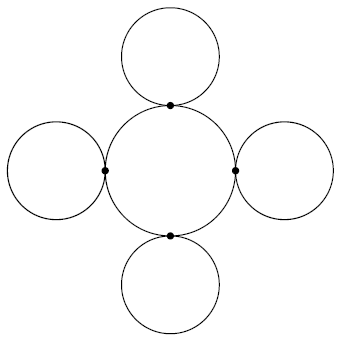}}_{24576}+\underbrace{\includegraphics[scale=0.12]{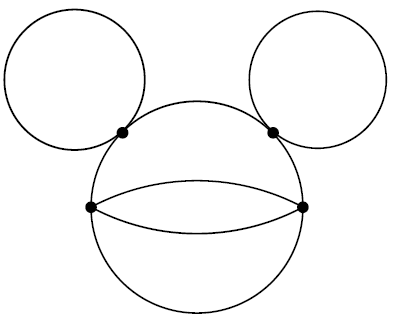}}_{49152}+\underbrace{\includegraphics[scale=0.12]{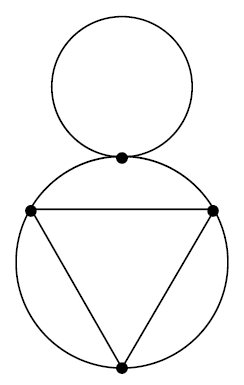}}_{98304}\label{FeynD}
 \end{align}

In Appendix \ref{II} we implement computationally this process. Note that the first two diagrams are disconnected. This process must be repeated for all the others possible RC-magic squares. At the end, after multiplying each Feynman graph subset by the respective factor (\ref{Prod}), we add the multiplicities of all the equivalent diagrams generated. If we divide by $(2m)!2^{2m}$ all the multiplicities we obtain the Feynman graphs multiplicities used in ref.\cite{Kleinert}. On the other hand, if we divide by $(4!)^{m}m!(2m)!2^{2m}$ we obtain for each diagram $1/s$, with $s$ the corresponding Feynman graph symmetry factor (see this multiplicative factors directly in (\ref{Der})).

\section{The permutation group and the set of RC-magic squares}\label{sec3}

We see that the total set of RC-magic squares with row and column sum equal to 4 provide all the vacuum Feynman graphs and the respective multiplicities in $\phi^{4}$ theory. As we will see later, the number $\mathcal{N}(m)$ of total $m\times m$ RC-magic squares grows rapidly with $m$. In particular for $m=1,2,3,4,5,6\cdots$ we have for $\mathcal{N}(m)$

\begin{equation}\label{seq}
1,5,120,10147,2224955,1047649905\cdots
\end{equation}
  these are many matrices. For example, at $4$-order we should apply the procedure used in (\ref{FeynD}) over the remaining $10146$ RC-magic squares. The sequence \ref{seq} is well known in the literature, see the OEIS sequence A172806, A257493 and the references therein.
  
  However, there are possible shortcuts. Let us show that for growing $m$ and for every $m\times m$ RC-magic square, there is a big number of equivalent RC-magic squares, which generates the same subset of vacuum Feynman graphs. This decreases considerably the number of RC-magic squares required to build all the vacuum Feynman graphs at $m$-order. 
  
  Suppose that one has performed a permutation of two different rows of a RC-magic square, evidently this is also a RC-magic square. In the $\aleph$ representation this is equivalent to interchange the two $\aleph$'s function. Only that this generates the same product of $\aleph$'s functions since the product is the usual (commutative multiplication). Now, suppose that two different columns are permutated, as we see the columns are associated with one variable $x_{i}$, so in principle the permutations change the $\aleph$'s functions that contain the $x_{i}$ and $x_{j}$ variables. At the end, when applying the distributive property in the new RC-magic square, the free propagators products terms are the same except that the variables $x_{i}$ and $x_{j}$ are interchanged. Thus, the column permutation correspond with one interchange of the vertices $x_{i}$ and $x_{j}$ in all the diagrams corresponding with the initial RC-magic square. Therefore columns and row permutation in a RC-magic square do not change the associated subset of vacuum Feynman graphs. It is evident that the multiplicity factor (\ref{Prod}) is identical for RC-magic squares related by a row or column permutation.
  
  Given an arbitrary matrix $\mathbf{A}$ of dimension $m\times m$, the possible $m!$ row permutations can be expressed by means of the left action $\mathbf{P}_{i}\cdot\mathbf{A}$, with $\mathbf{P}_{i}\in \mathcal{P}_{m\times m}$ and $\mathcal{P}_{m\times m}$ the set of $m\times m$ permutation matrices. The right action $\mathbf{A}\cdot \mathbf{P}_{i}$ induces the $m!$ possible column permutations. Thus, given an matrix $\mathbf{A}$ the set of different RC-magic squares obtained by row or column permutation $\mathbf{P}_{i}\cdot\mathbf{A}\cdot\mathbf{P}_{j}$ (including the matrix $\mathbf{A}$) contains at most $(m!)^{2}$ elements, (if there are matrices $\mathbf{P}_{a}$ and $\mathbf{P}_{b}$ such that $\mathbf{P}_{a}\cdot\mathbf{A}\cdot \mathbf{P}_{b}=\mathbf{A}$, this number is smaller). As we see, this different RC-magic squares generates identical sets of Feynman graphs.
  
  The row and column permutation group induces an equivalence relation in the set of RC-magic squares with equivalence classes of at most $(m!)^{2}$ elements. At the end, we choose an arbitrary element (RC-magic square) $\mathbf{A}_{i}$ in every equivalence class. Carrying out the procedure (\ref{FeynD1}) and (\ref{FeynD}) in each matrix $\mathbf{A}_{i}$, we obtain a set of Feynman diagrams $\mathcal{F}\left[\mathbf{A}_{i}\right]$ (with the respective generated multiplicities). The corresponding total multiplicities and all the possible diagrams at $m$-order are generated by
  
  \begin{equation}\label{Multi}
  \sum_{i} \mathfrak{N}_{\mathbf{A}_{i}}\times \prod_{j=1}^{m}\left[\frac{4!}{a_{j1}!a_{j2}!\cdots a_{jm}!}\right]_{i}\times \mathcal{F}\left[\mathbf{A}_{i}\right],
  \end{equation}   
  where $i$ indexes each equivalence class, $\mathfrak{N}_{\mathbf{A}_{i}}$ is the number of elements present in the equivalence class $i$, and $a_{jk}$ are the components of the chosen matrix $\mathbf{A}_{i}$. As discussed below, the number of equivalence classes, and therefore, the number of RC-magic squares required to generate the Feynman graphs for $m=1,2,3,4,5,6\cdots$ is respectively
  
  \begin{equation}
  1,3,9,43,264,2804\cdots
  \end{equation}    
  at least until 6-order, the number of equivalence classes match with the OEIS sequence A232216 generated in \cite{Geloun}. Given the total set of RC-magic squares for every order and the permutation matrices, the Burnside lemma in group theory \cite{Burnside} allows to find this sequence, Specifically, at $m$-order be $\mathcal{X}$ the set of $m\times m$ RC magic squares, and $\mathcal{G}$ the total group of row-column permutations acting in $\mathcal{X}$, this is, if $g(i,j) \in \mathcal{G}$ and $\mathbf{A}\in \mathcal{X}$, then $g(i,j)\cdot\mathbf{A}=\mathbf{P}_{i}\cdot\mathbf{A}\cdot\mathbf{P}_{j}$. Be $n(i,j)$ the number of elements $\mathbf{A}_{l}$ in $\mathcal{X}$ invariant by the row-column permutation $g(i,j)$, this is $g(i,j)\cdot\mathbf{A}_{l}=\mathbf{A}_{l}$, then the number of equivalence classes $n_{eq}$ in $\mathcal{X}$ induced by $\mathcal{G}$ is:
  
  \begin{equation}
  n_{eq}=\frac{1}{(m!)^{2}}\sum_{i=1}^{m!}\sum_{j=1}^{m!}n(i,j)
  \end{equation}
  for example, in $m=2$ we have only two permutation matrices, and five RC-magic squares, the number $n(i,j)$ are respectively $n(1,1)=5$, $n(1,2)=1$, $n(2,1)=1$ and $n(2,2)=5$. Then $n_{eq}=12/4=3$.
   
  As an example, we generate all the third-order Feynman graphs in $\phi^{4}$ theory. There are $9$ equivalence classes of $3\times 3$ RC-magic squares whose row and column sum is $4$. We choose the following nine representatives
  
  \begin{equation}\label{Third}
	\left(\begin{array}{ccc}
	4 & 0 & 0  \\
	0 & 4 & 0  \\
	0 & 0 & 4 
	\end{array}\right), 	
	\left(\begin{array}{ccc}
	4 & 0 & 0  \\
	0 & 3 & 1  \\
	0 & 1 & 3 
	\end{array}\right),
		\left(\begin{array}{ccc}
		4 & 0 & 0  \\
		0 & 2 & 2  \\
		0 & 2 & 2 
		\end{array}\right),
			\left(\begin{array}{ccc}
			3 & 1 & 0  \\
			0 & 1 & 3  \\
			1 & 2 & 1 
			\end{array}\right),
			\left(\begin{array}{ccc}
			3 & 1 & 0  \\
			0 & 3 & 1  \\
			1 & 0 & 3 
			\end{array}\right),
				\left(\begin{array}{ccc}
				3 & 1 & 0  \\
				0 & 2 & 2  \\
				1 & 1 & 2 
				\end{array}\right),
			\left(\begin{array}{ccc}
			2 & 2 & 0  \\
			0 & 2 & 2  \\
			2 & 0 & 2 
			\end{array}\right),	
				\left(\begin{array}{ccc}
				2 & 2 & 0  \\
				1 & 1 & 2  \\
				1 & 1 & 2 
				\end{array}\right),
					\left(\begin{array}{ccc}
					1 & 2 & 1  \\
					2 & 1 & 1  \\
					1 & 1 & 2 
					\end{array}\right)		
  \end{equation}
   
   with the number of elements $\mathfrak{N}_{3\times 3}$ $\{6,18,9,18,12,36,6,9,6\}$ in each equivalence class respectively, let us determine $\mathcal{F}[\mathbf{A}_{i}]$ for this nine matrices, in accordance with the procedure in (\ref{FeynD1}) and (\ref{FeynD}) for $m=3$.

    	\begin{align}
    		\left(\begin{array}{ccc}
    			4 & 0 & 0  \\
    			0 & 4 & 0  \\
    			0 & 0 & 4  
    		\end{array}\right)\to \underbrace{\includegraphics[scale=0.25]{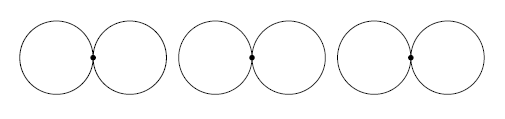}}_{13824}
    	\end{align}
    	
    		\begin{align}
    			\left(\begin{array}{ccc}
    				4 & 0 & 0  \\
    				0 & 3 & 1  \\
    				0 & 1 & 3  
    			\end{array}\right)\to \underbrace{\includegraphics[scale=0.25]{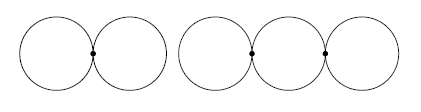}}_{13824}
    		\end{align}
    	
    		\begin{align}
    			\left(\begin{array}{ccc}
    				4 & 0 & 0  \\
    				0 & 2 & 2  \\
    				0 & 2 & 2  
    			\end{array}\right)\to \underbrace{\includegraphics[scale=0.25]{diag31.png}}_{1536}+\underbrace{\includegraphics[scale=0.25]{diag32.png}}_{6144}+\underbrace{\includegraphics[scale=0.25]{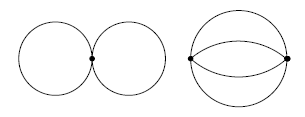}}_{6144}
    		\end{align}
    	
    	\begin{align}
    		\left(\begin{array}{ccc}
    			3 & 1 & 0  \\
    			0 & 1 & 3  \\
    			1 & 2 & 1  
    		\end{array}\right)\to \underbrace{\includegraphics[scale=0.25]{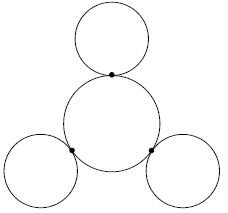}}_{4608}+\underbrace{\includegraphics[scale=0.25]{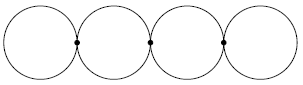}}_{9216}
    	\end{align}
    		\begin{align}
    			\left(\begin{array}{ccc}
    				3 & 1 & 0  \\
    				0 & 3 & 1  \\
    				1 & 0 & 3  
    			\end{array}\right)\to \underbrace{\includegraphics[scale=0.25]{diag34.png}}_{13824}
    		\end{align}	
    			\begin{align}
    				\left(\begin{array}{ccc}
    					3 & 1 & 0  \\
    					0 & 2 & 2  \\
    					1 & 1 & 2  
    				\end{array}\right)\to \underbrace{\includegraphics[scale=0.25]{diag32.png}}_{1536}+ \underbrace{\includegraphics[scale=0.25]{diag34.png}}_{3072}+ \underbrace{\includegraphics[scale=0.25]{diag35.png}}_{3072}+ \underbrace{\includegraphics[scale=0.25]{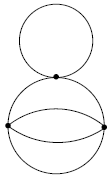}}_{6144}
    			\end{align}
    			\begin{align}
    				\left(\begin{array}{ccc}
    					2 & 2 & 0  \\
    					0 & 2 & 2  \\
    					2 & 0 & 2  
    				\end{array}\right)\to \underbrace{\includegraphics[scale=0.25]{diag31.png}}_{512}+\underbrace{\includegraphics[scale=0.25]{diag32.png}}_{3072}+\underbrace{\includegraphics[scale=0.25]{diag35.png}}_{6144}+\underbrace{\includegraphics[scale=0.25]{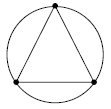}}_{4096}
    			\end{align}
    			\begin{align}
    				\left(\begin{array}{ccc}
    					2 & 2 & 0  \\
    					1 & 1 & 2  \\
    					1 & 1 & 2  
    				\end{array}\right)\to \underbrace{\includegraphics[scale=0.25]{diag32.png}}_{512}+\underbrace{\includegraphics[scale=0.25]{diag33.png}}_{1024}+\underbrace{\includegraphics[scale=0.25]{diag35.png}}_{2048}+\underbrace{\includegraphics[scale=0.25]{diag37.png}}_{4096}+\underbrace{\includegraphics[scale=0.25]{diag34.png}}_{2048}+\underbrace{\includegraphics[scale=0.25]{diag36.png}}_{4096}
    			\end{align}	
    			
    		\begin{align}
    			\left(\begin{array}{ccc}
    				1 & 2 & 1  \\
    				2 & 1 & 1  \\
    				1 & 1 & 2  
    			\end{array}\right)\to \underbrace{\includegraphics[scale=0.25]{diag34.png}}_{512}+\underbrace{\includegraphics[scale=0.25]{diag35.png}}_{3072}+\underbrace{\includegraphics[scale=0.25]{diag36.png}}_{6144}+\underbrace{\includegraphics[scale=0.25]{diag37.png}}_{4096}
    		\end{align}
    
   the size $\mathfrak{N}_{\mathbf{A}_{i}}\leq 36$ of each equivalence class is respectively $6,18,9,18,12,36,6,9,6$ and the product (\ref{Prod}) associated with each matrix (which is identical for all the matrices in the equivalence class) is respectively $1,16,36,192,64,288,216,864,1728$. Thus, applying the formula (\ref{Multi}) and adding the topological equivalent graphs we obtain the total multiplicitie $\mathcal{M}_{\mathrm{T}}$ of each third-order vacuum graph. If we divide $\mathcal{M}_{\mathrm{T}}$ by  $6!\times 2^{6}$ we obtain the multiplicities $\mathcal{M}_{\mathrm{K}}$ obtained by Kleinert in ref.\cite{Kleinert}, and dividing by $(4!)^{3}\times3!\times6!\times2^{6}$ we obtain $1/s$ with $s$ the corresponding symmetry factor of each graph. In particular

 \begin{equation}
   \underbrace{\includegraphics[scale=0.25]{diag31.png}}_{\mathcal{M}_{\mathrm{T}}=1244160, \,\,\,\, \mathcal{M}_{\mathrm{K}}=27, \,\,\,\, s=3072}
 \end{equation}
 
  \begin{equation}
  \underbrace{\includegraphics[scale=0.25]{diag32.png}}_{\mathcal{M}_{\mathrm{T}}=29859840, \,\,\,\, \mathcal{M}_{\mathrm{K}}=648, \,\,\,\, s=128}
  \end{equation}
  
  \begin{equation}
  \underbrace{\includegraphics[scale=0.25]{diag33.png}}_{\mathcal{M}_{\mathrm{T}}=9953280, \,\,\,\, \mathcal{M}_{\mathrm{K}}=216, \,\,\,\, s=384}
  \end{equation}

   \begin{equation}
   \underbrace{\includegraphics[scale=0.25]{diag34.png}}_{\mathcal{M}_{\mathrm{T}}=79626240, \,\,\,\, \mathcal{M}_{\mathrm{K}}=1728, \,\,\,\, s=48}
   \end{equation}
   
    \begin{equation}
    \underbrace{\includegraphics[scale=0.25]{diag35.png}}_{\mathcal{M}_{\mathrm{T}}=119439360, \,\,\,\, \mathcal{M}_{\mathrm{K}}=2592, \,\,\,\, s=32}
    \end{equation}
    
    \begin{equation}
    \underbrace{\includegraphics[scale=0.25]{diag36.png}}_{\mathcal{M}_{\mathrm{T}}=159252480, \,\,\,\, \mathcal{M}_{\mathrm{K}}=3456, \,\,\,\, s=24}
    \end{equation}
    
    \begin{equation}
    \underbrace{\includegraphics[scale=0.25]{diag37.png}}_{\mathcal{M}_{\mathrm{T}}=79626240, \,\,\,\, \mathcal{M}_{\mathrm{K}}=1728, \,\,\,\, s=48}
    \end{equation}

       \section{The set of RC-magic squares}\label{sec5} 
       
        In our construction, the calculus of the $m\times m$ RC-magic square representatives  is important to determine the set and multiplicities of all the $m$-order Feynman graphs. This calculation can be implemented by the following algorithm, which consists of two parts: the first one build a set $\mathcal{A}$ of matrices which contains at least one member of each equivalence class. The second one discard the equivalent elements and only preserve one element matrix of each equivalence class. At first sight, a minimal construction of the set $\mathcal{A}$ is not trivial, and in order to move forward we are forced to use a ``maximal" construction of the set $\mathcal{A}$ which build the matrices, testing all the possible constructions. In the $m\times m$ case such construction can be arduous, since when testing all the possibilities some must be discarded for not satisfying the equal row and column sum; what would be equivalent to analyzing a vast number of possibilities.
       
       Fortunately, the equivalence relation induced by the permutation group can be induced in any $n\times m$ set of matrices. In this case, the left action operation $\mathbf{P}_{i}\cdot \mathbf{A}$ is induced by the set $\mathcal{P}_{n\times n}$ of permutation matrices. Given a set of $n\times m$ matrix representatives of the total $n \times m$ matrix set with row sum equal to 4 and column sum less or equal to 4 and $n < m$, we can use the algorithm to calculate the matrices representatives in the same total set of $(n+1)\times m$ matrices. In this case, the set $\mathcal{A}=\mathcal{A}_{n+1}$ is constructed adding a new $n+1$ row to all the $n\times m$ matrices representatives and testing all the possibles $n+1$ rows such that the $(n+1)\times m$ matrix constructed have row sum equal to 4 and column sum less or equal to 4. The set  $\mathcal{A}_{n+1}$ is formed by all the $(n+1)\times m$ constructed matrices that satisfy such condition. This set contains at least one element of each equivalence class of the $(n+1)\times m$ matrix total set with the mentioned sum condition. Since we take all the possibles new row additions in each $n\times m$ matrix representative, this must be the case.
       
       The second part of the algorithm work in the following way, listing each element of $\mathcal{A}_{n+1}$ in a list, we take all the possible permutations for each matrix $\mathbf{A}_{i}$ in $\mathcal{A}_{n+1}$ this produce the set $\mathfrak{P}_{i}$. The set $\mathcal{G}_{i}=\mathfrak{P}_{i}\cap \mathcal{A}_{n+1}$ is formed by all the $(n+1)\times m$ matrices equivalent to $\mathbf{A}_{i}$ (including $\mathbf{A}_{i}$) contained in $\mathcal{A}_{n+1}$. Be $\mathfrak{H}_{i}$ the position of each element of  $\mathcal{G}_{i}$ in the list $\mathcal{A}_{n+1}$, $\mathfrak{H}_{i}$ is a set of different positive integers. For two equivalent matrices $\mathbf{A}_{i}$ and $\mathbf{A}_{j}$ we have $\mathfrak{H}_{i}=\mathfrak{H}_{j}$, instead if  $\mathbf{A}_{i}$ and $\mathbf{A}_{j}$ are not equivalent we have $\mathfrak{H}_{i}\cap\mathfrak{H}_{j}=\emptyset$. Taking all the different $\mathfrak{H}_{i}$ sets, we choose one element (for example the first) in every $\mathfrak{H}$. Taking the corresponding elements in the list $\mathcal{A}_{n+1}$, we obtain a matrix representative set of the $(n+1)\times m$ total set of matrices with row sum equal to 4 a column sum less or equal to 4.
       
       Thus, the $m\times m$ representatives can be obtained applying the algorithm $m-1$ times starting to the $1\times m$ matrix representatives. Obtained the set of $(m-1)\times m$ representatives exist a unique way to build each element of $\mathcal{A}_{m}$ which is a set of $m\times m$ RC-magic squares. Applying the second part of the algorithm to $\mathcal{A}_{m}$, we obtain the $m\times m$ RC-magic squares representatives. For $m\geq 4$ we have five $1\times m$ representatives which correspond with the sum partitions of 4

       	\begin{equation}
       	(4,0,\cdots,0),\,\,\,\,\,\, (3,1,0,\cdots,0),\,\,\,\,\,\, (2,2,0,\cdots,0),\,\,\,\,\,\, (2,1,1,0,\cdots,0),\,\,\,\,\,\, (1,1,1,1,0,\cdots,0),
       	\end{equation}
       
       if $m=2$ the $1\times 2$ representatives are $(4,0)$,$(3,1)$,$(2,2)$ and if $m=3$ we have $(4,0,0)$,$(3,1,0)$,$(2,2,0)$,$(2,1,1)$.	
       
       This algorithm, which code is implemented in APPENDIX.\ref{IV}, allows the calculation of the matrices representatives and the number of elements in each equivalence class. Anyhow, the algorithm realize left and right multiplications of matrices with the total set of permutations in the matricial representation (which, implements the column and row permutations). For each matrix, we have $(m!)^{2}$ products of matrices, thus for growing $m$ the computation represents a huge challenge, but a one that may have some ways of dealing with.
       
      The group of permutations not only induces equivalence classes in the total set of RC-magic squares, it has a deeper relationship with this set due to the Birkhoff-von Neumann theorem \cite{Brualdi}. This theorem establishes that any $m\times m$ doubly stochastic matrix $\mathbf{S}$ (matrices with real matrix-coefficients whose row and columns sums are equal to 1) is a finite linear combination of $m\times m$ permutations matrices in the following way 
      
       \begin{equation}\label{convex}
       \mathbf{S}=\lambda_{1}\mathbf{P}_{1}+\cdots \lambda_{k}\mathbf{P}_{k}, \,\,\,\,\,\,\,\, \sum_{j=1}^{k}\lambda_{j}=1
       \end{equation}
       with $\mathbf{P}_{j} \in \mathcal{P}_{m\times m}$ and $\lambda_{j} > 0$, the conditions in the linear coefficients $\lambda_{j}$ makes the linear combination a {\it convex} decomposition. An arbitrary RC-magic square with row and column sum equal to $d$ can be converted in a doubly stochastic matrix with rational coefficients if we multiply the RC magic square by the scalar $1/d$, this new matrix is doubly stochastic and have at least one convex descomposition. The general proof of the Birkhoff-von Neumann theorem implies that any RC-magic square $\mathbf{A}$ of sum $d$ can be written as a sum of $d$ (not necessarily distinct) permutation matrices. If only $k$ of this $d$ matrices are different, we can generalize (\ref{convex}) for the RC-magic square $\mathbf{A}$:
   
       \begin{equation}\label{convex2}
       \mathbf{A}=d_{1}\mathbf{P}_{1}+\cdots d_{k}\mathbf{P}_{k}, \,\,\,\,\,\,\,\, \sum_{j=1}^{k}d_{j}=d
       \end{equation}
       where $d_{j}$ are the times that the different $k$ permutation matrices appear, thus $1\geq d_{j}\geq d$. Abusing the language, we call (\ref{convex2}) a RC convex decomposition in the RC-magic square case. In particular, the only possible $d_{j}$'s in a RC convex decomposition for $d=4$ are $\{4\}$, $\{3,1\}$, $\{2,2\}$, $\{2,1,1\}$ and $\{1,1,1,1\}$. For some RC-magic squares, this decomposition is not unique. Apart from this possible repetition of the matrices, combining (\ref{convex2}) in all the possibles ways, we generate the total set of $m\times m$ RC-magic squares.
       
     We devise, below, a way to characterize the set of representatives in a more efficient way with the help of the Birkhoff-von Neumann theorem. As an example at third-order there are the following six permutations matrices
       
       \begin{align}
      \mathbf{P}_{1}= \left(\begin{array}{ccc}
       1 & 0 & 0 \\
       0 & 1 & 0 \\
       0 & 0 & 1
       \end{array}\right), \mathbf{P}_{2}= \left(\begin{array}{ccc}
       1 & 0 & 0 \\
       0 & 0 & 1 \\
       0 & 1 & 0
       \end{array}\right), \mathbf{P}_{3}= \left(\begin{array}{ccc}
       0 & 1 & 0 \\
       1 & 0 & 0 \\
       0 & 0 & 1
       \end{array}\right)\nonumber \\
      \mathbf{P}_{4} =\left(\begin{array}{ccc}
        	0 & 1 & 0 \\
        	0 & 0 & 1 \\
        	1 & 0 & 0
        \end{array}\right), \mathbf{P}_{5}= \left(\begin{array}{ccc}
        0 & 0 & 1 \\
        1 & 0 & 0 \\
        0 & 1 & 0
    \end{array}\right), \mathbf{P}_{6}= \left(\begin{array}{ccc}
    0 & 0 & 1 \\
    0 & 1 & 0 \\
    1 & 0 & 0
\end{array}\right)
       \end{align}  
       
       The RC convex decompositions of the 9 representative matrices (\ref{Third}) are
       
       \begin{equation}\label{1}
       	\left(\begin{array}{ccc}
       	4 & 0 & 0  \\
       	0 & 4 & 0  \\
       	0 & 0 & 4 
       	\end{array}\right)= 4\mathbf{P}_{1}
       \end{equation}
       
       \begin{equation}\label{2}
       \left(\begin{array}{ccc}
        4 & 0 & 0  \\
        0 & 3 & 1  \\
        0 & 1 & 3 
       \end{array}\right)=3\mathbf{P}_{1}+\mathbf{P}_{2}
       \end{equation}
       
       \begin{equation}\label{3}
       \left(\begin{array}{ccc}
       4 & 0 & 0  \\
       0 & 2 & 2  \\
       0 & 2 & 2 
       \end{array}\right)=2\mathbf{P}_{1}+2\mathbf{P}_{2}
       \end{equation}
       
       \begin{equation}\label{4}
       \left(\begin{array}{ccc}
       3 & 1 & 0  \\
       0 & 1 & 3  \\
       1 & 2 & 1 
       \end{array}\right)= \mathbf{P}_{1}+ \mathbf{P}_{4}+2\mathbf{P}_{2}
       \end{equation}
       
       \begin{equation}\label{5}
       \left(\begin{array}{ccc}
       3 & 1 & 0  \\
       0 & 3 & 1  \\
       1 & 0 & 3 
       \end{array}\right)=3\mathbf{P}_{1}+\mathbf{P}_{4}
       \end{equation}
       
       \begin{equation}\label{6}
       	\left(\begin{array}{ccc}
       	3 & 1 & 0  \\
       	0 & 2 & 2  \\
       	1 & 1 & 2 
       	\end{array}\right)= 2\mathbf{P}_{1}+ \mathbf{P}_{4}+\mathbf{P}_{2}
       \end{equation}
       
       \begin{equation}\label{7}
       \left(\begin{array}{ccc}
       2 & 2 & 0  \\
       0 & 2 & 2  \\
       2 & 0 & 2 
       \end{array}\right)=2\mathbf{P}_{1}+2\mathbf{P}_{4}
       \end{equation}
       
       \begin{equation}\label{8}
       	\left(\begin{array}{ccc}
       	2 & 2 & 0  \\
       	1 & 1 & 2  \\
       	1 & 1 & 2 
       	\end{array}\right)= \mathbf{P}_{1}+\mathbf{P}_{2}+\mathbf{P}_{3}+\mathbf{P}_{4}
       \end{equation}
       
       the last matrix have two possible descompositions:

       	\begin{equation}\label{9}
       		\left(\begin{array}{ccc}
       		1 & 2 & 1  \\
       		2 & 1 & 1  \\
       		1 & 1 & 2 
       		\end{array}\right)=2\mathbf{P}_{3}+ \mathbf{P}_{6}+ \mathbf{P}_{2}= \mathbf{P}_{1}+ \mathbf{P}_{3}+ \mathbf{P}_{4}+ \mathbf{P}_{5}
       	\end{equation}

      If we take all the subsets of $\mathcal{P}_{3\times 3}$ with $1,2,3$ and 4 elements respectively, we see that the elements in the subsets have a property in the non zero components that characterize in an unambiguous way each representative:
      \begin{itemize}
      \item The one-element subsets characterize completely the equivalence class represented by (\ref{1}), in this equation any matrix $\mathbf{P}_{j}$ establish one valid representative.
      \item The two-elements subsets can be divided in two parts with two properties: the two elements have one identical non zero component or do not have any non zero common component (from now on, when referring to common components will be understood that this are non zero components). The two permutation matrices in (\ref{2}) and (\ref{3}) have one component in common, instead the two permutation matrices in $\ref{5}$ and $\ref{7}$ do not have any common component. If we change this two permutation matrices by another satisfying the same property, we obtain an equivalent RC-magic square (that is to say another RC-magic square in the same equivalence class). The interchange of the $\lambda$'s only produces one permutation in the RC-magic square.
     \item  The three-elements subsets can be divided in two parts: The three permutations matrices have common components (in particular, one permutation matrix have two different components in common, one component with each one of the other matrices, and the other two matrices do not have any common component between them) or do not have any common component. The three permutation matrices in (\ref{4}) have the first property, with the fact that $\lambda=2$ is associated with the matrix that have two component in common with the other two. The three permutation matrices in (\ref{6}) also have the first property with the difference that $\lambda=2$ is associated with anyone of the two matrices that do not have common components between them. In (\ref{9}) the three permutations matrices do not have components in common. Here also the substitution of the permutation matrices by other with the same property produce an equivalent RC-magic square.
     
    \item At last, the four element subsets can be divided in two parts corresponding with two different properties in the number of common components (here the properties are in the number of common elements). The permutations matrices in (\ref{8}) and (\ref{9}) have this properties respectively and the substitution by other matrices with the same properties produce equivalent RC-magic squares.   
    \end{itemize}
    
    This construction can be generalized to larger orders. Particularly, it allows to build a technique to know if two arbitrary magic squares are not equivalent. Suppose that the arbitrary $m \times m$ RC-magic square $\mathbf{A}$ have $n$ different RC convex decompositions, each one with a particular common component property between the corresponding permutation matrices. Any RC-magic square $\mathbf{A}^{\prime}$ equivalent to $\mathbf{A}$ satisfy $\mathbf{A}^{\prime}=\mathbf{P}_{a}\cdot\mathbf{A}\cdot\mathbf{P}_{b}$, with $\mathbf{P}_{a}$ and $\mathbf{P}_{a}$ $\in$ $\mathcal{P}_{m\times m}$. Thereby, for a particular RC convex decomposition we have
    
\begin{equation}\label{Perm}
\mathbf{A}^{\prime}=\mathbf{P}_{a}\cdot\left(\sum_{j=1}^{k}\lambda_{j}\mathbf{P}_{j}\right)\cdot \mathbf{P}_{b}=\sum_{j=1}^{k}\lambda_{j}\mathbf{P}_{a}\cdot\mathbf{P}_{j}\cdot\mathbf{P}_{b}= \sum_{j=1}^{k}\lambda_{j}\mathbf{P}_{j}^{\prime}
\end{equation}

Since the permutation operation is identical for all the permutation matrices $\mathbf{P}_{j}$, (\ref{Perm}) will be one RC convex decomposition for $\mathbf{A}^{\prime}$ with the same common component property that the RC convex decomposition $\sum_{j=1}^{k}\lambda_{j}\mathbf{P}_{j}$ for $\mathbf{A}$. Since we take one arbitrary RC convex decomposition, the same is valid for all the other $n-1$ decompositions. 

Suppose now that $\mathbf{A}^{\prime}$ have other different RC convex decomposition in addition to the $n$ RC convex decompositions founded. Since $\mathbf{A}=\mathbf{P}_{a}^{-1}\cdot\mathbf{A}^{\prime}\cdot\mathbf{P}_{b}^{-1}$, this implies that $\mathbf{A}$ have $n+1$ RC convex decompositions. This is a contradiction since by hypothesis $\mathbf{A}$ have only $n$ different decompositions. Thus, defining $\mathcal{D}_{\mathbf{A}}$ and $\mathcal{D}_{\mathbf{B}}$ as the sets that contains all the RC convex decomposition of $\mathbf{A}$ and $\mathbf{B}$ respectively, we have the following result
\begin{itemize}
\item If the RC-magic square $\mathbf{A}$ and $\mathbf{B}$ are equivalent, then the two RC convex decompositions sets $\mathcal{D}_{\mathbf{A}}$ and $\mathcal{D}_{\mathbf{B}}$ have the same number of elements and an one to one correspondence between $\mathcal{D}_{\mathbf{A}}$ and $\mathcal{D}_{\mathbf{B}}$ such that the decompositions related with the correspondence have the same common component property. 
\end{itemize}

The contrapositive equivalent formulation of this result is more useful:

\begin{itemize}
\item If the sets $\mathcal{D}_{\mathbf{A}}$ and $\mathcal{D}_{\mathbf{B}}$ have different number the elements or if the elements of this sets can not be matched in an one to one correspondence with the  common component property, then $\mathbf{A}$ and $\mathbf{B}$ are not equivalent. 
\end{itemize}

Thus, we have one useful and simple criterion to check if two arbitrary RC-magic squares are not equivalent. For third and fourth-order, is possible list all the possible RC convex decomposition testing all the possible sums of permutation matrices, the complete set of RC convex decomposition of an arbitrary RC magic square is founded looking in the list the different RC convex decompositions that generate the arbitrary RC magic square. For larger orders, the known Hall theorem in combinatorial theory \cite{Diestel} offers a way to find the complete set of RC convex decomposition of an arbitrary RC-magic square. The idea is simple, every $m\times m$ RC-magic square of sum $d$ can be represented by a bipartite graph with $2m$ vertices, one vertex partition is represented by the $m$ rows and the other vertex partition by the $m$ columns. The coefficient $a_{ij}$ is the number of edges that conect the vertex $i$ with the vertex $j$. There are a total of $md$ edges, in particular for every RC magic square, exist at least one form to associate the total $md$ edges in $d$ perfect matchings of the $2m$ vertices (every perfect matching is one bipartite graph with $2m$ vertices and $m$ edges, which represent one $m\times m$ permutation matrix), this association correspond with an unique RC-convex decomposition. Therefore, The different associations of the $md$ edges in $d$ perfect matching, correspond with the different RC-convex decomposition.

 The converse of this criterion is not true, that is to say, if the sets  $\mathcal{D}_{\mathbf{A}}$ and $\mathcal{D}_{\mathbf{B}}$ have the same number of elements and an one to one correspondence that maps each element of $\mathcal{D}_{\mathbf{A}}$ in an respective element of $\mathcal{D}_{\mathbf{B}}$ with the same common component property, then the matrices $\mathbf{A}$ and $\mathbf{B}$ are not necessarily equivalent. As an example, consider the following two $4\times 4$ RC-magic squares with an unique RC convex decomposition

	\begin{equation}\label{41}
	\left(\begin{array}{cccc}
	2 & 1 & 1 & 0 \\
	0 & 0 & 2 & 2 \\
	0 & 2 & 0 & 2 \\
	2 & 1 & 1 & 0
	\end{array}\right)= \left(\begin{array}{cccc}
	0 & 0 & 1 & 0 \\
	0 & 0 & 0 & 1 \\
	0 & 1 & 0 & 0 \\
	1 & 0 & 0 & 0
	\end{array}\right)+\left(\begin{array}{cccc}
	0 & 1 & 0 & 0 \\
	0 & 0 & 1 & 0 \\
	0 & 0 & 0 & 1 \\
	1 & 0 & 0 & 0
	\end{array}\right)+\left(\begin{array}{cccc}
	1 & 0 & 0 & 0 \\
	0 & 0 & 0 & 1 \\
	0 & 1 & 0 & 0 \\
	0 & 0 & 1 & 0
	\end{array}\right)+\left(\begin{array}{cccc}
	1 & 0 & 0 & 0 \\
	0 & 0 & 1 & 0 \\
	0 & 0 & 0 & 1 \\
	0 & 1 & 0 & 0
	\end{array}\right)
	\end{equation} 
	and  
	
	\begin{equation}\label{42}
	\left(\begin{array}{cccc}
	2 & 1 & 1 & 0 \\
	0 & 1 & 1 & 2 \\
	0 & 2 & 2 & 0 \\
	2 & 0 & 0 & 2
	\end{array}\right)= \left(\begin{array}{cccc}
	0 & 0 & 1 & 0 \\
	0 & 0 & 0 & 1 \\
	0 & 1 & 0 & 0 \\
	1 & 0 & 0 & 0
	\end{array}\right)+\left(\begin{array}{cccc}
	0 & 1 & 0 & 0 \\
	0 & 0 & 0 & 1 \\
	0 & 0 & 1 & 0 \\
	1 & 0 & 0 & 0
	\end{array}\right)+\left(\begin{array}{cccc}
	1 & 0 & 0 & 0 \\
	0 & 0 & 1 & 0 \\
	0 & 1 & 0 & 0 \\
	0 & 0 & 0 & 1
	\end{array}\right)+\left(\begin{array}{cccc}
	1 & 0 & 0 & 0 \\
	0 & 1 & 0 & 0 \\
	0 & 0 & 1 & 0 \\
	0 & 0 & 0 & 1
	\end{array}\right)
	\end{equation}

     The two RC convex decompositions in each equation (\ref{41}) and (\ref{42}) have the same common component property, and each RC-magic square an unique RC convex decomposition. But this matrices are not equivalent. 

     \section{Discussion and conclusion}\label{sec6}
    
    The construction of Feynman graphs in quantum field theory is essentially a combinatorial problem. The question of how many ways exist to construct Feynman graphs given an number of vertices, edges and explicit rules (incidence numbers in graph theoretical language) for connecting vertices and edges, refers to traditional combinatorics problems which can be thought independently of any physical application. In combinatorics it is usual to find equivalences and relations between apparently different problems, in this work we find one combinatorial relation between counting vacuum Feynman graphs in $\phi^4$ theory, and the counting of a specific type of combinatorial matrices whose columns and rows coefficients sum 4, which in turn defines one combinatorial distribution problem (the $4m$ distribution ball in sec.\ref{sec2}). From a mathematical perspective, this relation can be understood as an application of combinatorial matrix theory in graph theory.
    
    From a physical perspective, two questions arise: does the combinatorial approach of this work simplify the calculation of the multiplicities? (or the symmetry factor) and, does the specific combinatorial character shown here have some direct and physical implication manifested in the integrals represented by the Feynman graphs?
    
    The first question is related to the importance in knowing the multiplicities for non-perturbative calculations in theories where the multiplicity combinatorics is non trivial, (for example $\phi^{4}$ theory). In this cases, the multiplicities appears explicitly in series of Feynman graphs, which could have some physical content. Resummation techniques are used to extract the physical content of this series, which tend to be divergent. This process has a deep relationship with the renormalization group algebra \cite{Broadhurst} (in the MB case, resummation techniques are used, see for example ref.\cite{Pavlyukh2}. In this case the zero dimensional approach is sufficient since, for the Feynman graphs in non-relativistic electrodynamics, the multiplicity combinatorics is trivial \cite{Castro}). In scalar theory, for non perturbative calculations, the multiplicity combinatorics must be considered in the process of resummation (for example, the didactic ref.\cite{Rivasseau} face this problem from another perspective). Only algorithmic and recursive calculations are know, this work transfer the algorithmic calculus in Feynman diagrams to the set of RC-magic squares (which are more tractable objects). There are many RC-magic squares, and we show that the relation with the permutation group decreases the number of RC-magic squares required for construct the Feynman graphs. In appendix \ref{IV} we implement a code for the algorithm described at the beginning of sec.\ref{sec5} for the calculation of the $m\times m$ RC-magic squares representatives and the size $\mathfrak{N}_{i}$ of each equivalence class, which works with all the elements of the permutation group $\mathcal{P}_{m\times m}$. The appendix \ref{II} contains a second algorithm for the calculation of the correspoding Feynman graphs multiplicities at $m$-order, which use the matrix representatives and the size of each equivalence class in the corresponding order. In principle, The first algorithm is independent of the Feynman graphs formulation, the second one is based in the relations of these distinct objects established in this work. We consider that an efficient computation of the RC-magic squares representatives and of the number $\mathfrak{N}_{i}$ would simplify considerably the construction of the Feynman graphs and the respective multiplicities. The Birkhoff-von Neumann theorem, could be the beginning of a more efficient algorithm  for the calculation of the RC-magic squares representatives. We believe that the present work offer an initial insight for such algorithmic construction. The second algorithm could also be improved, since we determine the topological equivalence between diagrams testing all the possibilities. For particular cases (see the appendices below), we show the execution times of our two codes, what could be useful for comparison with other existing algorithms \cite{Kleinert}. However, we consider that the two algorithms are not in their most efficient form, since they must test all the possible products with the permutation matrices. Our main objective here was to verify the validity of our result for orders larger than two or three. We hope that our work serves as a basis for the construction of more efficient algorithms. This being the case, computational complexity studies would be of general interest.
    
    The second question merits further study. We hope to investigate this in the near future. It is obvious that our construction can be easily generalised for any $\phi^N$ theory for $N \geq 3$ integer. The philosophy of this work can also be useful for any other perturbative quantum field theory, since the generating functional and the derivative functional construction process are also present.
   
    \section*{ACKNOWLEDGMENTS}

The authors thanks the Brazilian agency CNPq (Conselho Nacional de Desenvolvimento Científico e Tecnológico) for partial financial support.

    \appendix  
    
    \section{Computation of the permutation matrices, the RC-magic squares representatives and the size of the equivalence classes}\label{IV}
    
   The algorithm shown in sec.\ref{sec5} to calculate the RC-magic squares is valid for the permutation matrices (since this are RC-magic squares). In particular the filtering process for calculate the representatives of each equivalence class is unnecessary, in this respect all the permutation matrices are inequivalent. We begin with the $1\times m$ matrices, as example we calculate the permutation matrices for fifth-order. Thus, we have five $1\times 5$ matrices ($(1,0,0,0,0),(0,1,0,0,0),\cdots,(0,0,0,0,1)$). For the $2\times 5$ matrices we define
    	
    	\begin{lstlisting}
    	a[b1_, b2_, b3_, b4_, b5_] := 
    	DeleteCases[
    	Flatten[Table[
    	MatrixForm[({{b1, b2, b3, b4, b5}, {a1, a2, a3, a4, a5}})*
    	KroneckerDelta[a1 + a2 + a3 + a4 + a5, 1]*
    	HeavisideTheta[-(a1 + b1) + 1.1]*
    	HeavisideTheta[-(a2 + b2) + 1.1]*
    	HeavisideTheta[-(a3 + b3) + 1.1]*
    	HeavisideTheta[-(a4 + b4) + 1.1]*
    	HeavisideTheta[-(a5 + b5) + 1.1]], {a1, 0, 1}, {a2, 0, 
    		1 - a1}, {a3, 0, 1 - a1 - a2}, {a4, 0, 1 - a1 - a2 - a3}, {a5, 0,
    		1 - a1 - a2 - a3 - a4}]], 
    	MatrixForm[{{0, 0, 0, 0, 0}, {0, 0, 0, 0, 0}}]]
    	\end{lstlisting}
    	
    	The $2\times 5$ matrices are calculated using the 5 matrices $1\times 5$
    	
    	\begin{lstlisting}
        a25 = Union[a[1, 0, 0, 0, 0], a[0, 1, 0, 0, 0], a[0, 0, 1, 0, 0], 
    	a[0, 0, 0, 1, 0], a[0, 0, 0, 0, 1]]
    	\end{lstlisting}
    	in StandardForm
    	
    	\begin{lstlisting}
    	p25 = Table[a25[[n]][[1]],{n,1,Length[a25]}]
    	\end{lstlisting}
    	
    	Repeating the procedure, define
    	
    	\begin{lstlisting}
    	a[b1_, b2_, b3_, b4_, b5_, c1_, c2_, c3_, c4_, c5_] := 
    	DeleteCases[
    	Flatten[Table[
    	MatrixForm[({{b1, b2, b3, b4, b5}, {c1, c2, c3, c4, c5}, {a1, a2, 
    			a3, a4, a5}})*KroneckerDelta[a1 + a2 + a3 + a4 + a5, 1]*
    	HeavisideTheta[-(a1 + b1 + c1) + 1.1]*
    	HeavisideTheta[-(a2 + b2 + c2) + 1.1]*
    	HeavisideTheta[-(a3 + b3 + c3) + 1.1]*
    	HeavisideTheta[-(a4 + b4 + c4) + 1.1]*
    	HeavisideTheta[-(a5 + b5 + c5) + 1.1]], {a1, 0, 1}, {a2, 0, 
    		1 - a1}, {a3, 0, 1 - a1 - a2}, {a4, 0, 1 - a1 - a2 - a3}, {a5, 0,
    		1 - a1 - a2 - a3 - a4}]], 
    	MatrixForm[{{0, 0, 0, 0, 0}, {0, 0, 0, 0, 0}, {0, 0, 0, 0, 0}}]]
    	\end{lstlisting}
    	
    	and the $3\times 5$ matrices are given by
    	
    	\begin{lstlisting}
    	a35 = Union[Flatten[
    	Table[a[Flatten[p25[[n]]][[1]], Flatten[p25[[n]]][[2]], 
    	Flatten[p25[[n]]][[3]], Flatten[p25[[n]]][[4]], 
    	Flatten[p25[[n]]][[5]], Flatten[p25[[n]]][[6]], 
    	Flatten[p25[[n]]][[7]], Flatten[p25[[n]]][[8]], 
    	Flatten[p25[[n]]][[9]], Flatten[p25[[n]]][[10]]], {n, 1, 
    		Length[p25]}]]]
    	\end{lstlisting}
    	
    	\begin{lstlisting}
    	p35	= Table[a35[[n]][[1]],{n,1,Length[a35]}]
    	\end{lstlisting}
    	
    	we repeat this process, for the $4\times 5$ matrices we use $15$ variables in a[$\cdots$], and for the $5 \times 5$ matrices 20 variables. a55 contain the 5! permutation matrices in MatrixForm. For calculations is necesary to express a55 in the StandardForm, which is achieved with
    	
    	\begin{lstlisting}
    	P5=Table[a55[[m]][[1]],{m, 1, Length[a55]}]
    	\end{lstlisting}
    	
    	The generation times of the permutation matrices for 4, 5 and 6-order is fast, less than a second, using a conventional notebook.
    	
    	\subsection{The RC-magic squares representatives}

    Until now we saw that finding one representative matrix $\mathbf{A}_{i}$ in each equivalence class, and finding the size $\mathfrak{N}_{\mathbf{A}_{i}}$ of each equivalence class determine the $m$-order vacuum Feynman graphs (see this in (\ref{Multi})). Here, we implement the straightforward algorithm mentioned at the beginning of sec.\ref{sec5} for the determination of $\mathbf{A}_{i}$ and $\mathfrak{N}_{\mathbf{A}_{i}}$ using the program MATHEMATICA \cite{Mathematica}, which works, in principle, for all orders. The code builds up the RC-magic squares row by row, beginning with five $1\times m$ matrices for $m\geq 4$ (for $m<4$ we have a lower number of $1\times m$ matrices). For $5$-order this matrices are $(4,0,0,0,0)$,$(3,1,0,0,0)$,$(2,2,0,0,0)$,$(2,1,1,0,0)$, $(1,1,1,1,0)$. To obtain the matrices $2\times 5$ we define
    
    	\begin{lstlisting}
    	A[b1_, b2_, b3_, b4_, b5_] := 
    	DeleteCases[
    	Flatten[Table[
    	MatrixForm[({{b1, b2, b3, b4, b5}, {a1, a2, a3, a4, a5}})*
    	KroneckerDelta[a1 + a2 + a3 + a4 + a5, 4]*
    	HeavisideTheta[-(a1 + b1) + 4.1]*
    	HeavisideTheta[-(a2 + b2) + 4.1]*
    	HeavisideTheta[-(a3 + b3) + 4.1]*
    	HeavisideTheta[-(a4 + b4) + 4.1]*
    	HeavisideTheta[-(a5 + b5) + 4.1]], {a1, 0, 4}, {a2, 0, 
    	4 - a1}, {a3, 0, 4 - a1 - a2}, {a4, 0, 4 - a1 - a2 - a3}, {a5, 0,
    	4 - a1 - a2 - a3 - a4}]], 
    	MatrixForm[{{0, 0, 0, 0, 0}, {0, 0, 0, 0, 0}}]]
    	\end{lstlisting}

    Note that the function $\mathrm{A}$ have five variables, which correspond with the components of the five $1\times 5$ matrices. Thus, we define the set
    
    	\begin{lstlisting}
    	A2=Union[A[4,0,0,0,0],A[3,1,0,0,0],A[2,2,0,0,0],A[2,1,1,0,0],A[1,1,1,1,0]]
    	\end{lstlisting}
    and the set
    
    	\begin{lstlisting}
    	B2=Table[A2[[m]][[1]],{m,1,Length[A2]}]
    	\end{lstlisting}
   
       $\mathrm{A2}$ is the form appropriate in which MATHEMATICA interpret correctly the matrix elements of the set. While $\mathrm{B2}$ is the appropiate form for multiply matrices. Defining the permutation matrices sets $\mathcal{P}_{2\times 2}$, $\mathcal{P}_{3\times 3}$, $\mathcal{P}_{4\times 4}$ and $\mathcal{P}_{5\times 5}$ in StandardForm as P2, P3, P4 and P5 respectively (see the beginning of this Appendix) which have $2,6,24,120$ permutation matrices.  Finally we get the desired set of $2\times5$ matrices by
    	
    	\begin{lstlisting}
    	n = 1; While[n < h, H2[n] = A2[[1]];
    	G2 = Complement[A2, 
    	Intersection[A2, 
    	Flatten[Table[
    	MatrixForm[P2[[i]].B2[[1]].P5[[j]]], {i, 1, Length[P2]}, {j, 1, Length[P5]}]]]];
    	A2 = G2;
    	B2 = Table[A2[[m]][[1]], {m, 1, Length[A2]}]; n++]
    	\end{lstlisting}

     This code is a looping, which determine the set of $2\times 5$ representatives matrices denoted by the function H2[$\cdots$]. We see that the lists A2 and B2 are redefined in the process and the number h is such that, at the end of the process, we obtain A2$=$B2$=\emptyset$. Be $h_{min}$ the minimum of such numbers, So $h_{min}-1$ corresponds with the number of $2\times 5$ representatives matrices. In this case we have $h_{min}=45$, if $h<h_{min}$ (what is equivalent to saying A2$\neq\emptyset\neq$B2) we must repeat the process from the initial A2 and B2. In StandardForm we denote the set of  $2\times 5$ representatives as
     
       \begin{lstlisting}
        rc25=Table[H2[n][[1]],{n,1,hm}]
       \end{lstlisting}
     with hm$=h_{min}-1$. Therefore, we have a total of 44 matrices $2\times 5$. To get the matrices $3\times 5$ we repeat the procedure, the function $\mathrm{A}$ will have now $10$ variables
    
       \begin{lstlisting}
    	A[b1_, b2_, b3_, b4_, b5_, c1_, c2_, c3_, c4_, c5_] := 
    	DeleteCases[
    	Flatten[Table[
    	MatrixForm[({{b1, b2, b3, b4, b5}, {c1, c2, c3, c4, c5}, {a1, a2, 
    	a3, a4, a5}})*KroneckerDelta[a1 + a2 + a3 + a4 + a5, 4]*
    	HeavisideTheta[-(a1 + b1 + c1) + 4.1]*
    	HeavisideTheta[-(a2 + b2 + c2) + 4.1]*
    	HeavisideTheta[-(a3 + b3 + c3) + 4.1]*
    	HeavisideTheta[-(a4 + b4 + c4) + 4.1]*
    	HeavisideTheta[-(a5 + b5 + c5) + 4.1]], {a1, 0, 4}, {a2, 0, 
    	4 - a1}, {a3, 0, 4 - a1 - a2}, {a4, 0, 4 - a1 - a2 - a3}, {a5, 0,
    	4 - a1 - a2 - a3 - a4}]], 
    	MatrixForm[{{0, 0, 0, 0, 0}, {0, 0, 0, 0, 0}, {0, 0, 0, 0, 0}}]]
    	\end{lstlisting}

         in the A3 set we use the rc25 set

     	\begin{lstlisting}
    	A3 = Union[
    	Flatten[Table[
    	A[Flatten[rc25[[n]]][[1]], Flatten[rc25[[n]]][[2]], 
    	Flatten[rc25[[n]]][[3]], Flatten[rc25[[n]]][[4]], 
    	Flatten[rc25[[n]]][[5]], Flatten[rc25[[n]]][[6]], 
    	Flatten[rc25[[n]]][[7]], Flatten[rc25[[n]]][[8]], 
    	Flatten[rc25[[n]]][[9]], Flatten[rc25[[n]]][[10]]], {n, 1, 
    	Length[rc25]}]]]
    	\end{lstlisting}
    	in StandardForm
    	
    	\begin{lstlisting}
    	B3=Table[A3[[m]][[1]],{m,1,Length[A3]}]
    	\end{lstlisting}
    	
        Applying the last command
    
    \begin{lstlisting}
        n = 1; While[n < h, H3[n] = A3[[1]];
        G3 = Complement[A3, 
        Intersection[A3, 
        Flatten[Table[
        MatrixForm[P3[[i]].B3[[1]].P5[[j]]], {i, 1, Length[P3]}, {j, 1, Length[P5]}]]]];
        A3 = G3;
        B3 = Table[A3[[m]][[1]], {m, 1, Length[A3]}]; n++]
    \end{lstlisting}
    where $h_{min}=315$. The rc35 set is
    
    \begin{lstlisting}
        rc35=Table[H3[n][[1]],{n,1,hm}]
    \end{lstlisting}
    with hm$=h_{min}-1=314$
    
    We obtain 314 matrices $3\times 5$. Repeating the procedure again for the $4\times 5$ matrices we obtain $1021$ representatives. The 264 RC-magic squares of dimension $5\times 5$ representatives are obtained repeating the procedure one more time. In standardForm
    
    \begin{lstlisting}
        RC5=Table[H5[n][[1]],{n, 1, 264}]
    \end{lstlisting}
    
    The process is generalisable for generic $m$-order.
    
    To determine the size of each equivalence class size we define

    	\begin{lstlisting}
    	N5=Table[Length[DeleteDuplicates[
    	Flatten[Table[
    	MatrixForm[P5[[i]].RC5[[m]].P5[[j]]], {i, 1, 5!}, {j, 1, 5!}]]]]
    	,{m, 1, Length[RC5]}]
    	\end{lstlisting}
    
     The sequence (\ref{seq}) also can be obtained summing all the elements of the set Nm for each $m$ order. For example, in  $5$-order
    
    	\begin{lstlisting}
         Sum[N5[[i]],{i,1,264}]
    	\end{lstlisting}
    
    is equal to $2224955$.

The generation times of the representatives RC magic squares and the size of each equivalence class for 4, 5, and 6-order are approximately 1 second, 2 minutes and 18 hours respectively, using a conventional notebook.

    \section{Computation of the Feynman vacuum graphs multiplicities}\label{II}
    
      We will now to write the multiplicities of all the connected Feynman vacuum graphs for orders four and five. We use the program MATHEMATICA to perform explicitly the multiplicative process (\ref{FeynD1}). As we see, for each RC-magic square representative, we get one sum of $E$ functions, whose multiplicative coefficient are the multiplicities and the argument the adjacency matrix of a graph; at the end we use (\ref{Multi}) and we get a single set of adjacency matrices with the respective total multiplicities. Listing the multiplicities and the adjacency matrices in two list in such a way that each adjacency matrix and the corresponding multiplicity be indexed by the same positive integer. We add up the multiplicities of all the topologically equivalent graphs. In particular, if two graphs $\mathcal{G}_{1}$ and $\mathcal{G}_{2}$ of order $m$ are equivalent, then the corresponding $m\times m$ adjacency matrices $\mathbf{A}_{\mathcal{G}_{1}}$ and $\mathbf{A}_{\mathcal{G}_{2}}$ are connected by an row-column permutation of the kind
      
      \begin{equation}
      \mathbf{A}_{\mathcal{G}_{2}}=  \mathbf{P}_{a}\cdot\mathbf{A}_{\mathcal{G}_{1}}\cdot\mathbf{P}_{b}
      \end{equation}
      with $\mathbf{P}_{a}$ and $\mathbf{P}_{b}$ two permutation matrices in $\mathcal{P}_{m\times m}$ such that $\mathbf{P}_{a}\cdot \mathbf{P}_{b}= \mathbf{I}$ with $\mathbf{I}$ the $m\times m$ identity. Given $\mathbf{A}_{\mathcal{G}}$, how do calculate the equivalent diagrams? We list all the different products $\mathbf{P}_{i}\cdot\mathbf{A}_{\mathcal{G}}\cdot\mathbf{P}_{i}^{-1}$, and intersects with the previous adjacency matrix list. This gives all the adjacency matrices of the equivalent graphs, included $\mathbf{A}_{\mathcal{G}}$. The MATHEMATICA function Position[$\cdots$] determine the position of all this matrices in the adjacency matrix list. Using this information in the multiplicity list, (since each adjacency matrix and the associated multiplicity are indexed by the same number in the two lists) we add all the multiplicities corresponding to $\mathcal{G}$ giving the total multiplicity. 
     
      \subsection{Implementation of the algorithm for obtain the different Feynman graphs and the respectives multiplicities}
      
      Once given the representatives RC-magic squares, the size of each equivalence class and the permutation matrices at $m$-order; we write the MATHEMATICA code used for the calculus of the multiplicities. For fifth-order, we denote the respectives MATHEMATICA lists as RC5, N5 and P5. RC5 and N5 are indexed by the same natural number, this means that RC5[[m]] and N5[[m]] correspond with the equivalence class indexed by m.
      
      Define the set I5 of inverse matrices to P5
      
       \begin{lstlisting}
       I5=Table[Inverse[P5[[m]]], {m, 1, Length[P5]}]
       \end{lstlisting}
       where I5 and P5 are indexed by the same index. Now let's implement the multiplicative process of (\ref{FeynD1}), for this we write each one of the possible $\aleph$'s function in the $E$ format defined in (\ref{Mult}). We use only five functions, denoted by AL[4,0,0,0,0], AL[3,1,0,0,0], AL[2,2,0,0,0], AL[2,1,1,0,0] and AL[1,1,1,1,0]. The other variations of AL[4,0,0,0,0] are calculated from
       
        \begin{lstlisting}
        a1 = DeleteDuplicates[Table[P5[[m]].{4, 0, 0, 0, 0}, {m, 1, Length[P5]}]]
        
        a2 = DeleteDuplicates[Table[24 Exp[MatrixForm[P5[[m]].( {
        {2, 0, 0, 0, 0},
        {0, 0, 0, 0, 0},
        {0, 0, 0, 0, 0},
        {0, 0, 0, 0, 0},
        {0, 0, 0, 0, 0}
        } ).I5[[m]]]], {m, 1, Length[P5]}]]
        
        n = 1; While[n < Length[a1]+1, 
        AL[Flatten[a1[[n]]][[1]], Flatten[a1[[n]]][[2]], Flatten[a1[[n]]][[3]], 
        Flatten[a1[[n]]][[4]], Flatten[a1[[n]]][[5]]] = a2[[n]];
        n++]
        \end{lstlisting}
        
        The variations of AL[3,1,0,0,0]:
        
        \begin{lstlisting}
        a1 =  DeleteDuplicates[Table[P5[[m]].{3, 1, 0, 0, 0}, {m, 1, Length[P5]}]]
        
        a2 = DeleteDuplicates[Table[24 Exp[MatrixForm[P5[[m]].( {
        {1, 1, 0, 0, 0},
        {1, 0, 0, 0, 0},
        {0, 0, 0, 0, 0},
        {0, 0, 0, 0, 0},
        {0, 0, 0, 0, 0}
        } ).I5[[m]]]], {m, 1, Length[P5]}]]
        
        n = 1; While[n < Length[a1]+1, 
        AL[Flatten[a1[[n]]][[1]], Flatten[a1[[n]]][[2]], Flatten[a1[[n]]][[3]], 
        Flatten[a1[[n]]][[4]], Flatten[a1[[n]]][[5]]] = a2[[n]];
        n++]
        \end{lstlisting}
        
        The variations of AL[2,2,0,0,0]:
        
        \begin{lstlisting}
        a1 =  DeleteDuplicates[Table[P5[[m]].{2, 2, 0, 0, 0}, {m, 1, Length[P5]}]]
        
        a2 = DeleteDuplicates[Table[8 Exp[MatrixForm[P5[[m]].( {
        {1, 0, 0, 0, 0},
        {0, 1, 0, 0, 0},
        {0, 0, 0, 0, 0},
        {0, 0, 0, 0, 0},
        {0, 0, 0, 0, 0}
        } ).I5[[m]]]] + 16 Exp[MatrixForm[P5[[m]].( {
        {0, 2, 0, 0, 0},
        {2, 0, 0, 0, 0},
        {0, 0, 0, 0, 0},
        {0, 0, 0, 0, 0},
        {0, 0, 0, 0, 0}
        } ).I5[[m]]]], {m, 1, Length[P5]}]]
        
        n = 1; While[n < Length[a1]+1, 
        AL[Flatten[a1[[n]]][[1]], Flatten[a1[[n]]][[2]], Flatten[a1[[n]]][[3]], 
        Flatten[a1[[n]]][[4]], Flatten[a1[[n]]][[5]]] = a2[[n]];
        n++]
        \end{lstlisting}
      
       The variations of AL[2,1,1,0,0]:
       
       \begin{lstlisting}
       a1 =  DeleteDuplicates[Table[P5[[m]].{2, 1, 1, 0, 0}, {m, 1, Length[P5]}]]
       
       a2 = DeleteDuplicates[Table[8 Exp[MatrixForm[P5[[m]].( {
       {1, 0, 0, 0, 0},
       {0, 0, 1, 0, 0},
       {0, 1, 0, 0, 0},
       {0, 0, 0, 0, 0},
       {0, 0, 0, 0, 0}
       } ).I5[[m]]]] + 16 Exp[MatrixForm[P5[[m]].( {
       {0, 1, 1, 0, 0},
       {1, 0, 0, 0, 0},
       {1, 0, 0, 0, 0},
       {0, 0, 0, 0, 0},
       {0, 0, 0, 0, 0}
       } ).I5[[m]]]], {m, 1, Length[P5]}]]
       
       n = 1; While[n < Length[a1]+1, 
       AL[Flatten[a1[[n]]][[1]], Flatten[a1[[n]]][[2]], Flatten[a1[[n]]][[3]], 
       Flatten[a1[[n]]][[4]], Flatten[a1[[n]]][[5]]] = a2[[n]];
       n++]
       \end{lstlisting}
       
       The variations of AL[1,1,1,1,0]:
       
        \begin{lstlisting}
       a1 =  DeleteDuplicates[Table[P5[[m]].{1, 1, 1, 1, 0}, {m, 1, Length[P5]}]]
       
       a2 =DeleteDuplicates[Table[8 Exp[MatrixForm[P5[[m]].( {
       {0, 1, 0, 0, 0},
       {1, 0, 0, 0, 0},
       {0, 0, 0, 1, 0},
       {0, 0, 1, 0, 0},
       {0, 0, 0, 0, 0}
       } ).I5[[m]]]] + 8 Exp[MatrixForm[P5[[m]].( {
       {0, 0, 1, 0, 0},
       {0, 0, 0, 1, 0},
       {1, 0, 0, 0, 0},
       {0, 1, 0, 0, 0},
       {0, 0, 0, 0, 0}
       } ).I5[[m]]]] + 8 Exp[MatrixForm[P5[[m]].( {
       {0, 0, 0, 1, 0},
       {0, 0, 1, 0, 0},
       {0, 1, 0, 0, 0},
       {1, 0, 0, 0, 0},
       {0, 0, 0, 0, 0}
       } ).I5[[m]]]], {m, 1, Length[P5]}]]
       
       n = 1; While[n < Length[a1]+1, 
       AL[Flatten[a1[[n]]][[1]], Flatten[a1[[n]]][[2]], Flatten[a1[[n]]][[3]], 
       Flatten[a1[[n]]][[4]], Flatten[a1[[n]]][[5]]] = a2[[n]];
       n++]
       \end{lstlisting}
       
       This provides all the $\aleph$'s function in the $E$ format. Note that this process is easily generalizable for larger orders adding the necessary zero rows, and zero columns (at fourth-order we subtract the last zero row and the last zero column). Note that the $E$ function used in our code is the exponencial function Exp[$\cdots$], since the argument contains the MatrixForm[$\cdots$] function which is for MATHEMATICA an undefined object. This guarantees the property (\ref{Mult}) for all the possibilities.
       
       In order to represent the multiplicity factor (\ref{Prod}) we define
       
       \begin{lstlisting}
       M[m_] := 4!/(
       RC5[[m]][[1, 1]]!*RC5[[m]][[1, 2]]!*RC5[[m]][[1, 3]]!*
       RC5[[m]][[1, 4]]!*RC5[[m]][[1, 5]]!)*4!/(
       RC5[[m]][[2, 1]]!*RC5[[m]][[2, 2]]!*RC5[[m]][[2, 3]]!*
       RC5[[m]][[2, 4]]!*RC5[[m]][[2, 5]]!)*4!/(
       RC5[[m]][[3, 1]]!*RC5[[m]][[3, 2]]!*RC5[[m]][[3, 3]]!*
       RC5[[m]][[3, 4]]!*RC5[[m]][[3, 5]]!)*4!/(
       RC5[[m]][[4, 1]]!*RC5[[m]][[4, 2]]!*RC5[[m]][[4, 3]]!*
       RC5[[m]][[4, 4]]!*RC5[[m]][[4, 5]]!)*4!/(
       RC5[[m]][[5, 1]]!*RC5[[m]][[5, 2]]!*RC5[[m]][[5, 3]]!*
       RC5[[m]][[5, 4]]!*RC5[[m]][[5, 5]]!)
       \end{lstlisting}
       
       The distributive multiplication process (\ref{FeynD1}) is realized by
       
       \begin{lstlisting}
       A[m_] := Expand[
       M[m]*N5[[m]]*
       AL[RC5[[m]][[1, 1]], RC5[[m]][[1, 2]], RC5[[m]][[1, 3]], 
       RC5[[m]][[1, 4]], RC5[[m]][[1, 5]]]*
       AL[RC5[[m]][[2, 1]], RC5[[m]][[2, 2]], RC5[[m]][[2, 3]], 
       RC5[[m]][[2, 4]], RC5[[m]][[2, 5]]]*
       AL[RC5[[m]][[3, 1]], RC5[[m]][[3, 2]], RC5[[m]][[3, 3]], 
       RC5[[m]][[3, 4]], RC5[[m]][[3, 5]]]*
       AL[RC5[[m]][[4, 1]], RC5[[m]][[4, 2]], RC5[[m]][[4, 3]], 
       RC5[[m]][[4, 4]], RC5[[m]][[4, 5]]]*
       AL[RC5[[m]][[5, 1]], RC5[[m]][[5, 2]], RC5[[m]][[5, 3]], 
       RC5[[m]][[5, 4]], RC5[[m]][[5, 5]]]]
        \end{lstlisting}
        
        and the equivalent of (\ref{Multi}) is
        
        \begin{lstlisting}
     Z = Sum[A[m], {m, 1, Length[RC5]}]
       \end{lstlisting}
      which generate all the Feynman graphs with the respective multiplicities. The next pass is to determine which diagrams are equivalent, and add the multiplicities for all the equivalent diagrams.
      First, listing all the multiplicities associated with Z
      
       \begin{lstlisting}
      Mult = Table[Z[[m]][[1]], {m, 1, Length[Z]}]
       \end{lstlisting}
       
       Second, listing all the graphs (adjacency matrices) associated with Z
       
       \begin{lstlisting}
     G = Table[Z[[m]][[2, 2]], {m, 1, Length[Z]}]
      \end{lstlisting}
      
      Nevertheless, the elements of G are sums of matrices in the MatrixForm format, which must be added. For this, we define 
      
      \begin{lstlisting}
      L[m_, l_] := 
      If[Head[G[[m]][[l]]] === Times, 
      G[[m]][[l]][[1]]* G[[m]][[l]][[2, 1]], G[[m]][[l]][[1]]]
      
      Elem[m_] := If[Head[G[[m]]] === Plus, 
      Sum[L[m, l], {l, 1, Length[G[[m]]]}], G[[m]][[1]]*G[[m]][[2, 1]]]
      \end{lstlisting}
      
      Therefore, the adjacency matrices associated with Z are
      
      \begin{lstlisting}
     Graphs = Table[Elem[m], {m, 1, Length[G]}]
      \end{lstlisting}
      
      The two lists Graphs and Multi are indexed by the same natural numbers, the adjacency matrix Graphs[[$l$]] have multiplicity Multi[[$l$]]. In order to determine the total multiplicities of the different diagrams, we define first

      \begin{lstlisting}
      GCopy = Graphs
      \end{lstlisting}
      
      After, we obtain the different adjacency matrices and the corresponding multiplicities using
      
      \begin{lstlisting}
      n = 1; While[n < g, Adj[n] = MatrixForm[GCopy[[1]]];
      G2 = Intersection[Table[P5[[m]].GCopy[[1]].I5[[m]], {m, 1, Length[P5]}], GCopy];
      G3 = Complement[GCopy, G2];
      Deg[n] = 
      Flatten[Table[Flatten[Position[Graphs, G2[[m]]]], {m, 1, Length[G2]}]];
      GCopy = G3;
      n++]
      \end{lstlisting}
      
      This code is a looping, which determine the set of different adjacency matrices Adj[$\cdots$]. We see that the list GCopy is redefined in the process and the number $g$ is such that, at the end of the process, we obtain GCopy$=\emptyset$. Be $g_{min}$ the minimum of such numbers, So $g_{min}-1$ corresponds with the number of different adjacency matrices. At fifth-order we have $g_{min}=57$, if $g<g_{min}$ (what is equivalent to saying GCopy$\neq\emptyset$) we must repeat the process from 
      
      \begin{lstlisting}
      GCopy = Graphs
      \end{lstlisting}
      
      For a given adjacency matrix Adj[$m$], Deg[$m$] gives the equivalent diagrams in the list Graphs ($1\leq m < g_{min}$). Therefore, the multiplicity of Adj[$m$] is simply
      
      \begin{lstlisting}
      Sum[Mult[[n]], {n, Deg[m]}]
      \end{lstlisting}

   The different Feynman graphs of fifth\correct{-}order are given by
   
      \begin{lstlisting}
      Table[AdjacencyGraph[Adj[n][[1]]], {n, 1, gm}]
      \end{lstlisting}
  with gm=$g_{min}-1$.
  
  The generation times of the different Feynman graphs and the associated multiplicities for 4, 5 and 6-order are approximately 0.2 seconds, 15 seconds and 1 hour respectively using a conventional notebook.
    
     \subsection{All the fourth and fifth-order Feynman vacuum graphs
     	multiplicities}
      For disconnected graphs we verify for fourth and fifth-order the rule shown in \cite{Dong}. Particularly for a disconnected graph with $l$ connected components of which $r$ are different, we verify that the symmetry factor is
      
      \begin{equation}
      s_{d}=n_{1}!\cdots n_{r}!\times s_{1}^{n_{1}}\cdots s_{r}^{n_{r}}
      \end{equation}
      
      with $s_{i}$ the symmetry factor of the component $i$, $n_{i}$ the times it repeats and $n_{1}+\cdots+n_{r}=l \geq r$. Thus, we will only write the multiplicities for the connected Feynman graphs.
       
      \subsubsection{Fourth-order connected diagrams}
      
       \begin{equation}
       \underbrace{\includegraphics[scale=0.25]{diag41.png}}_{\mathcal{M}_{\mathrm{T}}=642105999360, \,\,\,\, \mathcal{M}_{\mathrm{K}}=62208, \,\,\,\, s=128}
       \end{equation}

       \begin{equation}
       \underbrace{\includegraphics[scale=0.20]{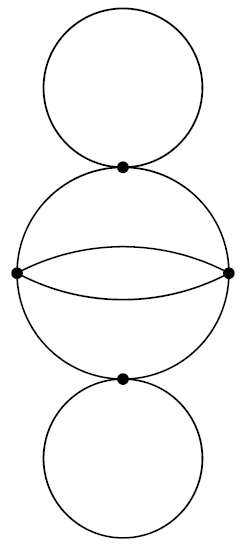}}_{\mathcal{M}_{\mathrm{T}}=2568423997440, \,\,\,\, \mathcal{M}_{\mathrm{K}}=248832, \,\,\,\, s=32}
       \end{equation}
       
       \begin{equation}
       \underbrace{\includegraphics[scale=0.20]{diag43.png}}_{\mathcal{M}_{\mathrm{T}}=1712282664960, \,\,\,\, \mathcal{M}_{\mathrm{K}}=165888, \,\,\,\, s=48}
       \end{equation}
       
       \begin{equation}
       \underbrace{\includegraphics[scale=0.20]{diag44.png}}_{\mathcal{M}_{\mathrm{T}}=1712282664960, \,\,\,\, \mathcal{M}_{\mathrm{K}}=165888, \,\,\,\, s=48}
       \end{equation}
       
       \begin{equation}
       \underbrace{\includegraphics[scale=0.20]{diag45.png}}_{\mathcal{M}_{\mathrm{T}}=2568423997440, \,\,\,\, \mathcal{M}_{\mathrm{K}}=248832, \,\,\,\, s=32}
       \end{equation}
       
       \begin{equation}
       \underbrace{\includegraphics[scale=0.20]{diag46.png}}_{\mathcal{M}_{\mathrm{T}}=5136847994880, \,\,\,\, \mathcal{M}_{\mathrm{K}}=497664, \,\,\,\, s=16}
       \end{equation}
       
       \begin{equation}
       \underbrace{\includegraphics[scale=0.20]{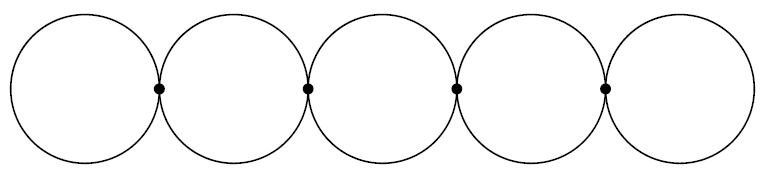}}_{\mathcal{M}_{\mathrm{T}}=1284211998720 , \,\,\,\, \mathcal{M}_{\mathrm{K}}=124416, \,\,\,\, s=64}
       \end{equation}
       
        \begin{equation}
        \underbrace{\includegraphics[scale=0.20]{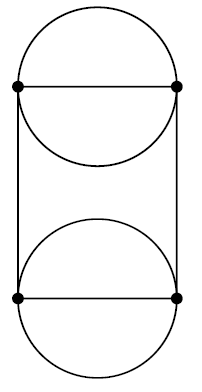}}_{\mathcal{M}_{\mathrm{T}}=570760888320 , \,\,\,\, \mathcal{M}_{\mathrm{K}}=55296, \,\,\,\, s=144}
        \end{equation}
        
        \begin{equation}
        \underbrace{\includegraphics[scale=0.20]{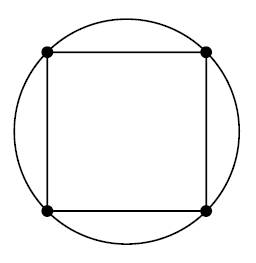}}_{\mathcal{M}_{\mathrm{T}}=642105999360 , \,\,\,\, \mathcal{M}_{\mathrm{K}}=62208, \,\,\,\, s=128}
        \end{equation}
        
         \begin{equation}
         \underbrace{\includegraphics[scale=0.20]{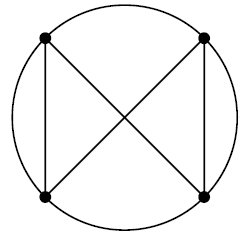}}_{\mathcal{M}_{\mathrm{T}}=2568423997440 , \,\,\,\, \mathcal{M}_{\mathrm{K}}=248832, \,\,\,\, s=32}
         \end{equation}
       
        \subsubsection{Fifth-order connected diagrams}
        
        \begin{equation}
        \underbrace{\includegraphics[scale=0.20]{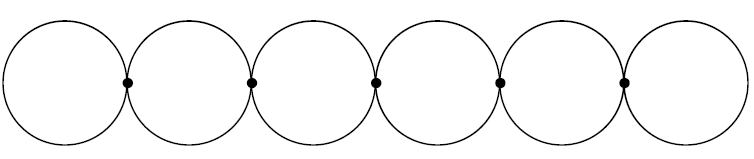}}_{\mathcal{M}_{\mathrm{T}}=27738979172352000 , \,\,\,\, \mathcal{M}_{\mathrm{K}}=7464960, \,\,\,\, s=128}
        \end{equation}
        
        \begin{equation}
        \underbrace{\includegraphics[scale=0.20]{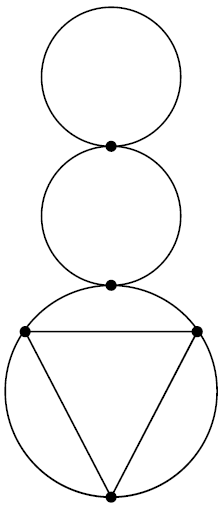}}_{\mathcal{M}_{\mathrm{T}}=110955916689408000 , \,\,\,\, \mathcal{M}_{\mathrm{K}}=29859840, \,\,\,\, s=32}
        \end{equation}
        
         \begin{equation}
         \underbrace{\includegraphics[scale=0.20]{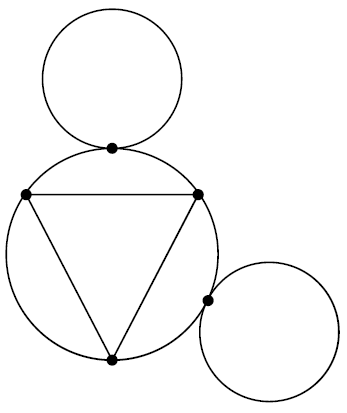}}_{\mathcal{M}_{\mathrm{T}}=221911833378816000 , \,\,\,\, \mathcal{M}_{\mathrm{K}}=59719680, \,\,\,\, s=16}
         \end{equation}
         
         \begin{equation}
         \underbrace{\includegraphics[scale=0.20]{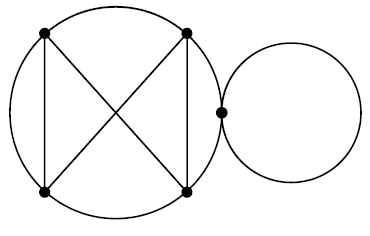}}_{\mathcal{M}_{\mathrm{T}}=221911833378816000 , \,\,\,\, \mathcal{M}_{\mathrm{K}}=59719680, \,\,\,\, s=16}
         \end{equation}
         
          \begin{equation}
          \underbrace{\includegraphics[scale=0.20]{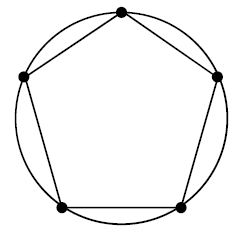}}_{\mathcal{M}_{\mathrm{T}}=11095591668940800 , \,\,\,\, \mathcal{M}_{\mathrm{K}}=2985984, \,\,\,\, s=320}
          \end{equation}
          
           \begin{equation}
           \underbrace{\includegraphics[scale=0.20]{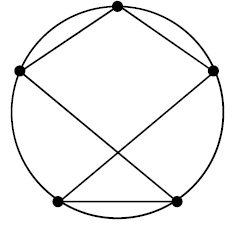}}_{\mathcal{M}_{\mathrm{T}}=110955916689408000 , \,\,\,\, \mathcal{M}_{\mathrm{K}}=29859840, \,\,\,\, s=32}
           \end{equation}
           
            \begin{equation}
            \underbrace{\includegraphics[scale=0.20]{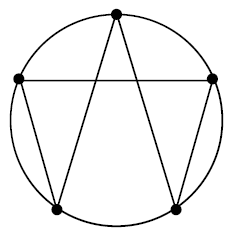}}_{\mathcal{M}_{\mathrm{T}}=221911833378816000 , \,\,\,\, \mathcal{M}_{\mathrm{K}}=59719680, \,\,\,\, s=16}
            \end{equation}
            
             \begin{equation}
             \underbrace{\includegraphics[scale=0.20]{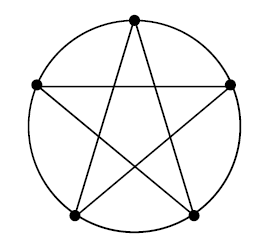}}_{\mathcal{M}_{\mathrm{T}}=29588244450508800 , \,\,\,\, \mathcal{M}_{\mathrm{K}}=7962624, \,\,\,\, s=120}
             \end{equation}

         \begin{equation}
         \underbrace{\includegraphics[scale=0.20]{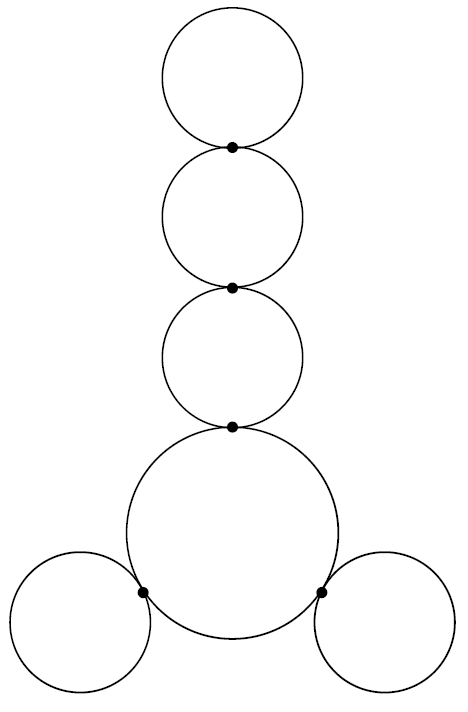}}_{\mathcal{M}_{\mathrm{T}}=55477958344704000 , \,\,\,\, \mathcal{M}_{\mathrm{K}}=14929920, \,\,\,\, s=64}
         \end{equation}
         
         \begin{equation}
         \underbrace{\includegraphics[scale=0.20]{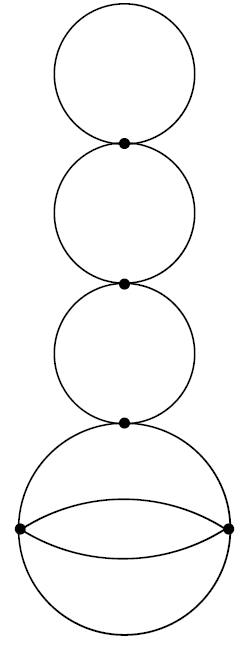}}_{\mathcal{M}_{\mathrm{T}}=36985305563136000 , \,\,\,\, \mathcal{M}_{\mathrm{K}}=9953280, \,\,\,\, s=96}
         \end{equation}
         
         \begin{equation}
         \underbrace{\includegraphics[scale=0.25]{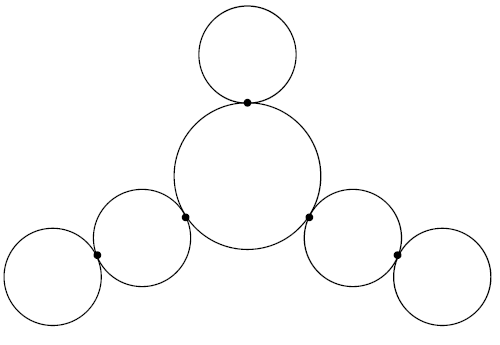}}_{\mathcal{M}_{\mathrm{T}}=55477958344704000 , \,\,\,\, \mathcal{M}_{\mathrm{K}}=14929920, \,\,\,\, s=64}
         \end{equation}
         
         \begin{equation}
         \underbrace{\includegraphics[scale=0.20]{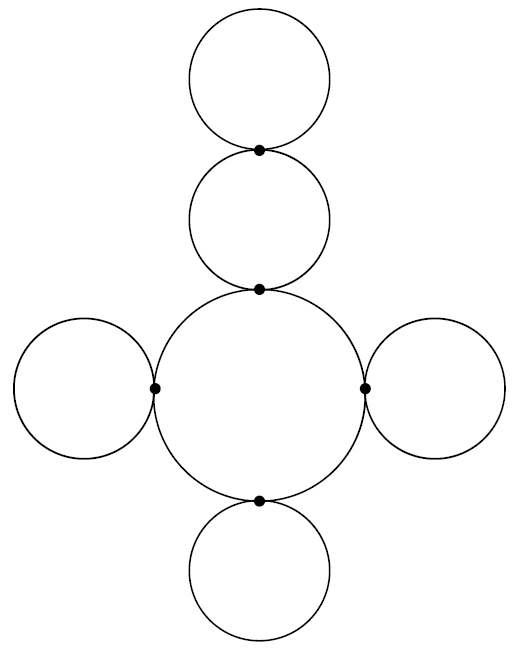}}_{\mathcal{M}_{\mathrm{T}}=55477958344704000 , \,\,\,\, \mathcal{M}_{\mathrm{K}}=14929920, \,\,\,\, s=64}
         \end{equation}
         
         \begin{equation}
         \underbrace{\includegraphics[scale=0.20]{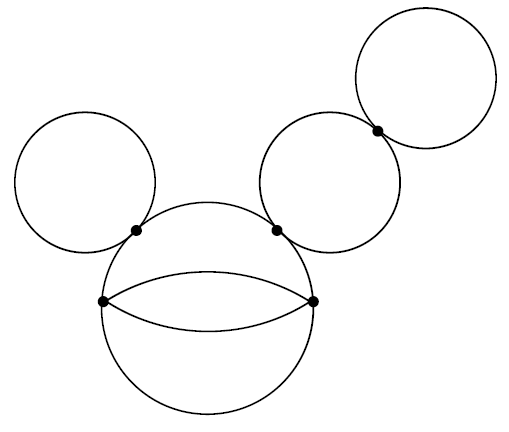}}_{\mathcal{M}_{\mathrm{T}}=73970611126272000 , \,\,\,\, \mathcal{M}_{\mathrm{K}}=19906560, \,\,\,\, s=48}
         \end{equation}
         
         \begin{equation}
         \underbrace{\includegraphics[scale=0.20]{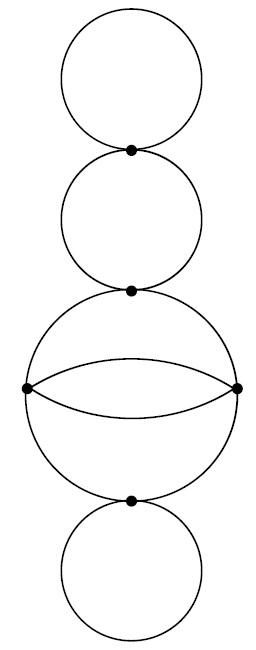}}_{\mathcal{M}_{\mathrm{T}}=110955916689408000 , \,\,\,\, \mathcal{M}_{\mathrm{K}}=29859840, \,\,\,\, s=32}
         \end{equation}
         
          \begin{equation}
          \underbrace{\includegraphics[scale=0.20]{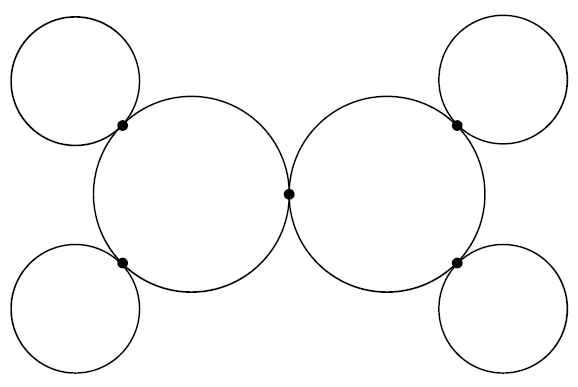}}_{\mathcal{M}_{\mathrm{T}}=27738979172352000 , \,\,\,\, \mathcal{M}_{\mathrm{K}}=7464960, \,\,\,\, s=128}
          \end{equation}
          
           \begin{equation}
           \underbrace{\includegraphics[scale=0.20]{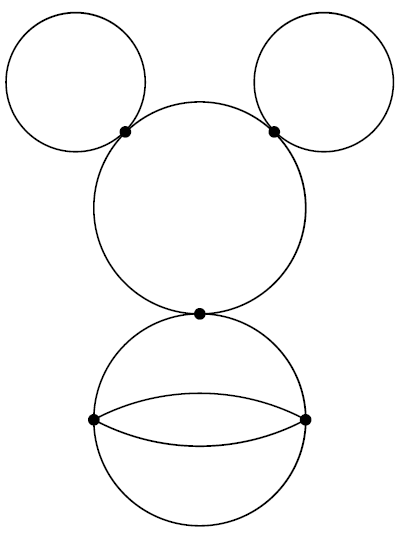}}_{\mathcal{M}_{\mathrm{T}}=36985305563136000 , \,\,\,\, \mathcal{M}_{\mathrm{K}}=9953280, \,\,\,\, s=96}
           \end{equation}
           
            \begin{equation}
            \underbrace{\includegraphics[scale=0.20]{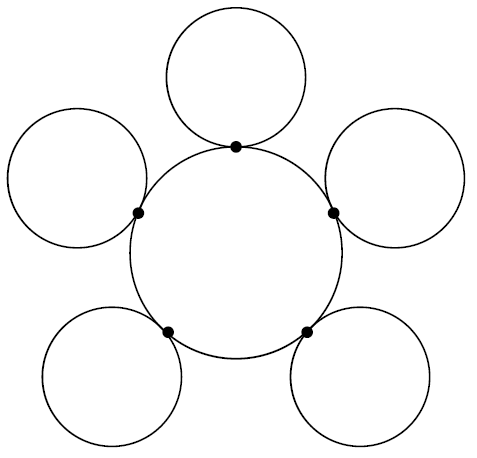}}_{\mathcal{M}_{\mathrm{T}}=11095591668940800 , \,\,\,\, \mathcal{M}_{\mathrm{K}}=2985984, \,\,\,\, s=320}
            \end{equation}
        
         \begin{equation}
         \underbrace{\includegraphics[scale=0.20]{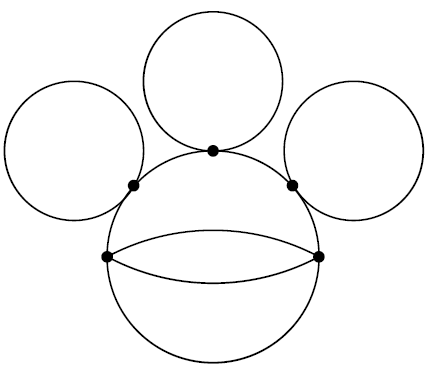}}_{\mathcal{M}_{\mathrm{T}}=36985305563136000 , \,\,\,\, \mathcal{M}_{\mathrm{K}}=9953280, \,\,\,\, s=96}
         \end{equation}
         
         \begin{equation}
         \underbrace{\includegraphics[scale=0.20]{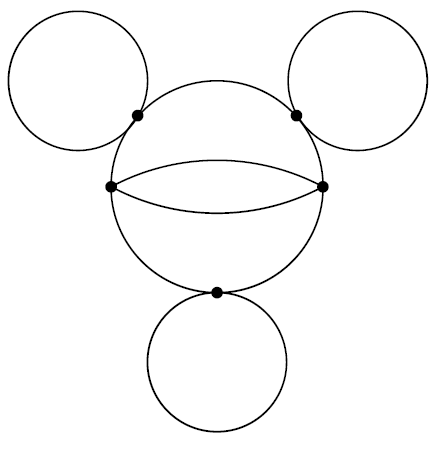}}_{\mathcal{M}_{\mathrm{T}}=110955916689408000 , \,\,\,\, \mathcal{M}_{\mathrm{K}}=29859840, \,\,\,\, s=32}
         \end{equation}
         
          \begin{equation}
          \underbrace{\includegraphics[scale=0.20]{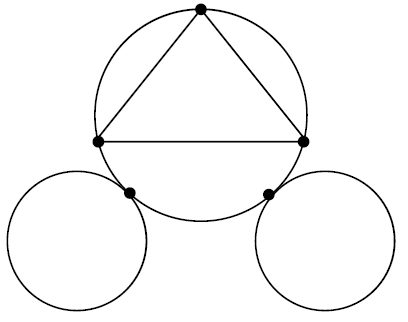}}_{\mathcal{M}_{\mathrm{T}}=110955916689408000 , \,\,\,\, \mathcal{M}_{\mathrm{K}}=29859840, \,\,\,\, s=32}
          \end{equation}
          
           \begin{equation}
           \underbrace{\includegraphics[scale=0.20]{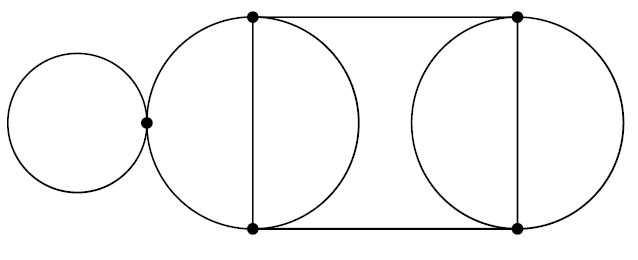}}_{\mathcal{M}_{\mathrm{T}}=73970611126272000 , \,\,\,\, \mathcal{M}_{\mathrm{K}}=19906560, \,\,\,\, s=48}
           \end{equation}
           
         \begin{equation}
         \underbrace{\includegraphics[scale=0.20]{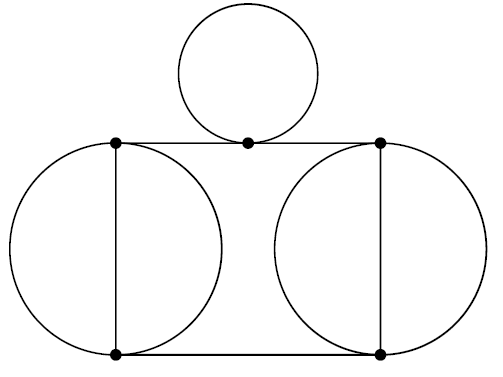}}_{\mathcal{M}_{\mathrm{T}}=24656870375424000 , \,\,\,\, \mathcal{M}_{\mathrm{K}}=6635520, \,\,\,\, s=144}
         \end{equation}   
         
         \begin{equation}
         \underbrace{\includegraphics[scale=0.20]{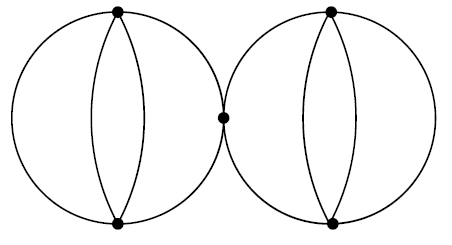}}_{\mathcal{M}_{\mathrm{T}}=12328435187712000 , \,\,\,\, \mathcal{M}_{\mathrm{K}}=3317760, \,\,\,\, s=288}
         \end{equation}
          
         \begin{equation}
         \underbrace{\includegraphics[scale=0.20]{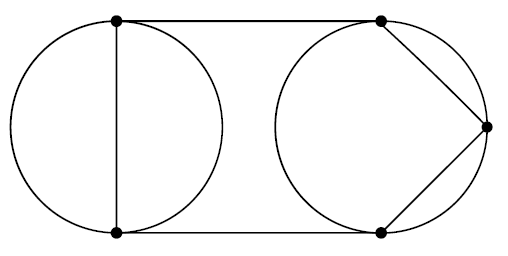}}_{\mathcal{M}_{\mathrm{T}}=73970611126272000 , \,\,\,\, \mathcal{M}_{\mathrm{K}}=19906560, \,\,\,\, s=48}
         \end{equation} 
         
          \begin{equation}
          \underbrace{\includegraphics[scale=0.20]{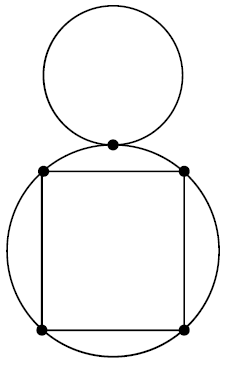}}_{\mathcal{M}_{\mathrm{T}}=110955916689408000 , \,\,\,\, \mathcal{M}_{\mathrm{K}}=29859840, \,\,\,\, s=32}
          \end{equation}
          
           \begin{equation}
           \underbrace{\includegraphics[scale=0.20]{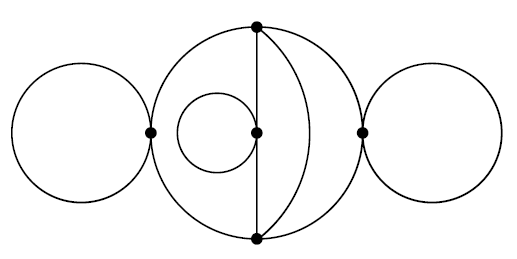}}_{\mathcal{M}_{\mathrm{T}}=36985305563136000 , \,\,\,\, \mathcal{M}_{\mathrm{K}}=9953280, \,\,\,\, s=96}
           \end{equation}
           
            \begin{equation}
            \underbrace{\includegraphics[scale=0.20]{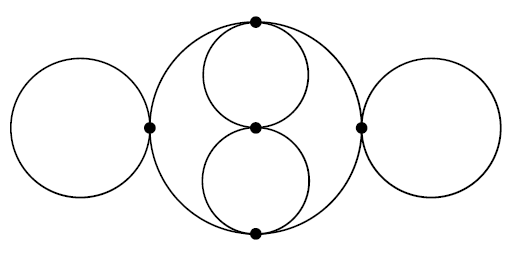}}_{\mathcal{M}_{\mathrm{T}}=55477958344704000 , \,\,\,\, \mathcal{M}_{\mathrm{K}}=14929920, \,\,\,\, s=64}
            \end{equation}
            
             \begin{equation}
             \underbrace{\includegraphics[scale=0.20]{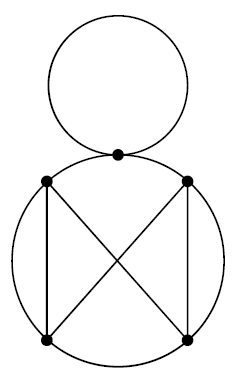}}_{\mathcal{M}_{\mathrm{T}}=221911833378816000 , \,\,\,\, \mathcal{M}_{\mathrm{K}}=59719680, \,\,\,\, s=16}
             \end{equation}
         \\
        For sixth order, we have 187 different Feynman graphs of which 97 are connected.
	
	\bibliography{RefsArticle}
	
\end{document}